\documentclass[aps,prd,twocolumn,nofootinbib,amsmath,amssymb,floatfix]{revtex4-2}

\usepackage{bm}
\usepackage{dsfont}
\usepackage{bbold}
\usepackage{amsthm}
\usepackage{graphicx}
\usepackage[colorlinks=true,linkcolor=blue,citecolor=blue,urlcolor=blue]{hyperref}

\newcommand{\dd}{\text{d}}
\newcommand{\Kc}{\mathcal{K}}
\newcommand{\Dc}{\mathcal{D}}
\newcommand{\Mellin}{\mathcal{M}}
\newcommand{\Id}{\mathbb{1}}
\newcommand{\Aff}{\mathrm{Aff}}
\newcommand{\GL}{GL}

\theoremstyle{plain}
\newtheorem{theorem}{Theorem}
\theoremstyle{definition}
\newtheorem{definition}{Definition}

\begin{document}

\title{Scheme transformations as the gauge group of DGLAP:\\ sum rules, classification and solution}
\author{Tommaso Rainaldi}
\email{tommaso.rainaldi@stonybrook.edu}
\affiliation{Department of Physics and Astronomy, Stony Brook University, New York 11794, USA}

\begin{abstract}
A parton density depends on the renormalization scheme, and any two
schemes are connected by a transformation that preserves the DGLAP
form of the evolution. In Mellin space such a transformation acts on
the evolution kernels as a gauge transformation, and we take this
point of view throughout the discussion. For the unpolarized densities we first ask
which transformations also preserve the charge-conjugation and flavor
symmetries of the evolution kernels and the momentum and valence
number sum rules. The symmetries reduce the transformation to a few
blocks, and the sum rules constrain only two Mellin moments, where
they leave a single freedom, the split of momentum between the quark
singlet and the gluon, and freeze the valence numbers. The resulting
classification is complete, because with asymptotic freedom as a
boundary condition a change of scheme is determined by its action on
the kernels. Inside this group, the transformations that preserve
positivity form a semigroup, and the same argument covers the
polarized densities. A transformation that fails only the sum rules
is not lost either, since a normalization that the sum rules
themselves determine repairs it, and we use this to define collinear
densities as integrals of transverse-momentum-dependent densities
without giving up the sum rules. We then use the same gauge freedom
to solve the evolution. Gauging the kernel away reduces DGLAP at any
order to a single second-order linear equation plus a quadrature, and
it organizes the estimate of missing higher orders in a way that
respects the sum rules exactly.
\end{abstract}

\maketitle

\makeatletter
\def\l@subsection#1#2{}
\makeatother
\tableofcontents

\section{Introduction}
\label{sec:intro}

A parton density is not a scheme-independent object. Its definition
requires a renormalization prescription and, while dimensional
regularization with modified minimal subtraction ($\overline{\rm MS}$) is
the standard choice, infinitely many other choices are allowed in
principle. It is then natural to ask which scheme choices are compatible
with the parton-model interpretation of the densities. In particular, we
would like the evolution kernels to keep their charge-conjugation and
flavor symmetry, the
partons to carry the hadron's momentum, and the valence content to be
fixed.

The freedom in this choice is as old as the next-to-leading order
calculations themselves
\cite{Altarelli:1978,Bardeen:1978,Altarelli:1979}, and the general
redefinition of coefficient functions and evolution kernels by a matrix
acting on flavor space was written down already in
Refs.~\cite{Curci:1980,Furmanski:1982}. The same papers contain,
in particular cases, the two constraints we study here. The Adler sum
rule was used in Refs.~\cite{Altarelli:1978,Altarelli:1979} to fix the
first moment of the quark kernel, so that the valence content survives
beyond leading order, and momentum conservation was imposed in
Ref.~\cite{Altarelli:1979} to fix the second moment of the otherwise
ambiguous DIS-scheme gluon. Scheme dependence as a source of theoretical
uncertainty was studied soon after
\cite{Politzer:1982,Stevenson:1986,Chyla:1989}, and one can even
eliminate the densities altogether in favor of scheme-invariant physical
evolution kernels \cite{Catani:1996,Blumlein:2004}.

Recently, the subject
has become topical again. Schemes designed for specific purposes, such
as the absence of quark-gluon mixing \cite{Oliveira:2013}, positivity of
the densities in $x$ space \cite{Candido:2020,Collins:2021,Candido:2023},
next-to-leading order matching to parton showers \cite{Jadach:2013},
and small-$x$ resummation, where the change from $\overline{\rm MS}$ to
the $Q_0\overline{\rm MS}$ scheme is an all-orders kernel
\cite{Bonvini:2016},
have been collected and compared systematically in
Refs.~\cite{Delorme:2025,Delorme:2026}, which also quantify how the
momentum and number sum rules respond to a change of scheme at order
$\alpha_s$. Sum rules have likewise been used to constrain the relation
between the factorization and renormalization scales
\cite{Rogers:2026}. A further case where the question arises is the
definition of a collinear density as the integral of its
transverse-momentum-dependent (TMD) counterpart over transverse
momentum \cite{Gonzalez-Hernandez:2023,Aslan:2024,delRio:2024,Rainaldi:2025efq},
which is a change of scheme in its own right.

This work is complementary to that program. However, instead of comparing given
schemes one at a time, we characterize the entire set of transformations
that preserve the symmetries and the sum rules exactly and at every
order. We work in Mellin space,
where scheme transformations become matrices and the DGLAP kernels
transform like gauge connections, and we impose the physical requirements
in order of their reach. The symmetries of the kernels act at every Mellin moment
and reduce the transformation, via Schur's lemma, to a small number of
blocks. The sum rules then act only at the first two moments and remove
most of the remaining freedom. The result is an explicit classification
with a transparent physical meaning, and we prove that it is complete.
Asymptotic freedom, used as a boundary condition, removes every residual
redundancy, so that nothing outside the classification survives. In the
second half of the paper we use the same structure to organize the
solution of the evolution. Since a one-dimensional connection is pure
gauge, solving DGLAP is itself a choice of scheme, and the entire
evolution, at any perturbative order, reduces to a single second-order
linear equation (a Riccati equation in disguise) plus a quadrature.

Many of the concepts and results collected here are known. The
transformation law of the kernels goes back to
Refs.~\cite{Curci:1980,Furmanski:1982}, the sum-rule conditions at order
$\alpha_s$ to Refs.~\cite{Altarelli:1978,Altarelli:1979}, and the
diagonalization of the singlet block, derivative term included, is used
in small-$x$ resummation, e.g. \cite{Bonvini:2016}, and closed analytic
solutions of the singlet evolution have been constructed by other
methods \cite{Simonelli:2024}. What this work adds is a unifying
language where the kernels are seen as gauge connections on flavor space, in which
these results become parts of a single construction. Within it, three
statements are, to our knowledge, new. The classification of
Eq.~\eqref{eq:final_group} is complete, because a change of scheme is
determined by its action on the kernels
(Theorems~\ref{thm:classification} and \ref{thm:injectivity}). The
uniqueness behind Theorem~\ref{thm:injectivity} is implicit in every
construction that solves for a transformation from its kernel, such as
the recursion of Appendix~\ref{sec:dressed}, and the theorem extends it
to the whole set of admissible transformations. The sum rules are never
an obstruction by themselves, since any transformation with the right
symmetries that fails them is repaired by a normalization fixed by a
few numbers. At one loop and for the momentum sum rule this is the
endpoint counterterm of Refs.~\cite{Delorme:2025,Delorme:2026}, and the
classification extends it to the number sum rules and to every order.
For collinear densities defined as integrals of TMDs it identifies the
scheme that keeps both the sum rules and the integral relation. And
the two parts combine into an analysis of the missing higher orders in the
evolution that separates what a given order knows exactly from what it
does not, and that phrases scheme variation as an uncertainty estimate
which respects the sum rules exactly and contains scale variation by a
fixed ratio as a special case.

The paper is divided into two parts that can be read independently.
Part~\ref{part:one} contains the classification.
In Section~\ref{sec:inputs} we collect the definitions and the physical
constraints, and formulate the scheme transformations in Mellin space.
In Section~\ref{sec:classification} we first impose the symmetries and then the
sum rules. Here, we derive the classification of the allowed schemes.
In Section~\ref{sec:injectivity} we prove that the classification is exhaustive
and explain why no further sum rules exist at other Mellin moments.
Section~\ref{sec:positivity} determines which of the classified
transformations preserve positivity of the densities in $x$ space, and
applies the answer to the evolution operator itself.
Section~\ref{sec:tmd} applies the classification to collinear densities
defined as integrals of TMDs.
Part~\ref{part:two} contains the solution. Section~\ref{sec:solution}
gauges the evolution away and reduces DGLAP to a single second-order
equation at any order. In Sec.~\ref{sec:mhou} we analyze the missing higher
orders, and in Sec.~\ref{sec:summary} we summarize both parts.
In Appendix~\ref{sec:soffer}, we briefly extend the positivity argument to polarized densities. Finally, in 
Appendix~\ref{sec:dressed} we relate the DGLAP Frobenius series solution to the well-known
dressed expansion of the evolution operator \cite{Vogt:2005}.

\part{Classification of the sum-rule-preserving scheme transformations}
\label{part:one}

In this part, we classify the DGLAP-preserving scheme transformations that also preserve the
charge-conjugation and flavor symmetries of the evolution kernels and
the sum rules of the parton densities. 

\section{Setup: densities, symmetries and sum rules}
\label{sec:inputs}

\subsection{Parton densities and DGLAP evolution}

We consider a theory with $N_f$ fermionic flavors, so that the parton
content of a hadron $h$ comprises $2N_f+1$ densities: quarks $q_i$,
antiquarks $\bar q_i$ ($i=1,\dots,N_f$), and the gluon $g$. A
renormalized density in a scheme $s$ is obtained from the bare one by a
multiplicative renormalization in the sense of Mellin convolution,
\begin{equation}
\begin{aligned}
    f^{(s)}_{i/h}(x;\mu)
    &= \int_x^1\frac{\dd\xi}{\xi}\,Z^{(s)}_{ij}(\xi;\mu)\,
      f_{0,j/h}\!\left(\frac{x}{\xi}\right)\\
    &\equiv \left(Z^{(s)}_{ij}\otimes f_{0,j/h}\right)(x;\mu),
\end{aligned}
    \label{eq:def_pdf}
\end{equation}
where $x$ is the light-cone momentum fraction and $\mu$ the
renormalization scale. A change of the factor $Z$ is, by definition, a
change of scheme. We restrict attention to schemes in which the densities
evolve in the DGLAP
form~\cite{Gribov:1972ri,Altarelli:1977zs,Dokshitzer:1977sg},
\begin{equation}
    \frac{\dd}{\dd\ln\mu^2}\,f^{(s)}_i(x;\mu)
    = \left(P^{(s)}_{ij}\otimes f^{(s)}_j\right)(x;\mu),
    \label{eq:DGLAP_x}
\end{equation}
with $P^{(s)}_{ij}(x;\mu)$ the matrix of splitting kernels in that
scheme. 

\subsection{The symmetries of the evolution kernels}
\label{sec:qcd_symmetries}

We work with mass-independent evolution kernels, as in
$\overline{\rm MS}$ and its relatives, so that the $N_f$ flavors enter
the evolution symmetrically. Charge conjugation invariance and flavor
symmetry then constrain the splitting matrix to the familiar form
\begin{equation}
\begin{gathered}
    P_{q_iq_j}=P_{\bar q_i\bar q_j}=\delta_{ij}P^{V}_{qq}+P^{S}_{qq},\\
    P_{q_i\bar q_j}=P_{\bar q_i q_j}=\delta_{ij}P^{V}_{q\bar q}+P^{S}_{q\bar q},\\
    P_{q_i g}=P_{qg},
    \qquad
    P_{g q_i}=P_{gq},
\end{gathered}
    \label{eq:kernel_symmetries}
\end{equation}
which leaves $7$ independent functions for $N_f>1$ (and $5$ for
$N_f=1$). 
Introducing the involution $C$ that exchanges
$q_i\leftrightarrow\bar q_i$ (leaving the gluon fixed) and the symmetric
group $S_{N_f}$ of permutations of the $N_f$ flavors, represented by
permutation matrices acting simultaneously on quark and antiquark
labels, Eq.~\eqref{eq:kernel_symmetries} is precisely the statement that
$P$ commutes with every element of the finite group
$G\equiv\langle C\rangle\times S_{N_f}$. The angle brackets denote the
group generated by an element, so $\langle C\rangle=\{\Id,C\}\cong
\mathbb{Z}_2$ since $C^2=\Id$, while $C$ by itself is just a matrix.
These constraints hold at every value of $x$ and, equivalently, at every
Mellin moment.

\subsection{The sum rules}
\label{sec:sum_rules_x}

Two further exact statements accompany Eq.~\eqref{eq:DGLAP_x} in any
scheme with a partonic interpretation. The momentum sum rule states that
the partons carry the hadron's momentum,
\begin{equation}
    \sum_{i}\int_0^1\dd x\; x\, f_i(x;\mu) = 1
    \qquad\text{for all }\mu ,
    \label{eq:momentum_sum_rule_x}
\end{equation}
and the $N_f$ valence number sum rules state that the net number of
quarks of each flavor is a fixed integer characteristic of the hadron
(e.g.\ $n_u=2$ and $n_d=1$ for the proton),
\begin{equation}
    \int_0^1\dd x\,\left[f_{q_i}(x;\mu) - f_{\bar q_i}(x;\mu)\right] = n_{q_i}
    \qquad\text{for all }\mu .
    \label{eq:number_sum_rule_x}
\end{equation}
Both statements are scale independent, so consistency with the evolution
imposes exact constraints on the kernels. Differentiating
Eq.~\eqref{eq:momentum_sum_rule_x} gives the familiar momentum
conservation of the splitting functions,
\begin{equation}
    \sum_{j}\int_0^1 \dd x\; x\, P_{ji}(x;\mu) = 0 \quad \forall\, i ,
    \label{eq:momentum_conservation_P}
\end{equation}
and similarly Eq.~\eqref{eq:number_sum_rule_x} requires the first moment
of the valence evolution kernel to vanish.

Each sum rule
constrains the kernels only at one specific moment of $x$ (the second and the
first, respectively), while the symmetry constraints hold at every
moment. This difference dictates the order of the analysis below. We
first impose the symmetries, which act everywhere, and then the sum
rules.

\subsection{Scheme transformations in Mellin space}
\label{sec:mellin}

Since every object above enters through Mellin convolutions, we transform
to moment space, where convolutions become simple matrix products
\begin{equation}
\begin{gathered}
    \overline{f}(z;\mu)
    \equiv \int_0^1\dd x\; x^{z-1} f(x;\mu),\\
    \gamma(z;\mu)\equiv\Mellin\{P\},
    \qquad
    \Kc(z;\mu)\equiv\Mellin\{K\},
\end{gathered}
\end{equation}
with $z\in\mathbb{C}$ the Mellin moment and $K(x;\mu)$ the $x$-space
kernel of a change of scheme. The Mellin variable is more
commonly denoted by $N$ but we use $z$ as a reminder that it is a complex
variable, since $N$ suggests an integer, and the analytic structure in
$z$ will matter repeatedly in what follows. A scheme transformation is then
an invertible $(2N_f+1)\times(2N_f+1)$ matrix acting on the densities,
\begin{equation}
    \overline{f}^{(s_2)}(z;\mu) = \Kc(z;\mu)\,\overline{f}^{(s_1)}(z;\mu).
    \label{eq:change_scheme}
\end{equation}
Demanding that the DGLAP form of the evolution survive the change of
scheme fixes the transformation law of the kernels, given at order
$\alpha_s$ in Ref.~\cite{Furmanski:1982} and in the all-orders form in,
e.g., Ref.~\cite{Delorme:2025}:
\begin{equation}
    \gamma^{(s_2)}
    = \Kc\,\gamma^{(s_1)}\Kc^{-1}
    + \left(\frac{\dd \Kc}{\dd\ln\mu^2}\right)\Kc^{-1} .
    \label{eq:gauge_transformation}
\end{equation}
This is formally the transformation of a gauge connection, with $\gamma$
playing the role of the gauge field and $\Kc$ that of the gauge
transformation. DGLAP evolution is then the parallel-transport equation
$\Dc_\gamma\overline f = 0$ for the covariant derivative
$\Dc_\gamma \equiv \dd/\dd\ln\mu^2-\gamma$. 

This analogy is not merely
cosmetic, since the inhomogeneous term of
Eq.~\eqref{eq:gauge_transformation} is what makes the classification
below nontrivial.

Let us now proceed by listing important properties of the scheme transformations that will be used repeatedly in what follows.

\textbf{(i) Group structure:} The composition of scheme changes obeys
$\Kc^{s_1\gets s_2}\Kc^{s_2\gets s_3}=\Kc^{s_1\gets s_3}$ with identity
and inverses, so at each fixed $z$ the transformations form a group
inside $\GL(2N_f+1,\mathbb{C})$.

\textbf{(ii) Boundary condition:} Asymptotic freedom demands that scheme
dependence disappear with the coupling:
\begin{equation}
    \Kc(z;\mu) = \Id + O(a_s) \quad\text{as } \mu\to\infty .
    \label{eq:asymptotic_identity}
\end{equation}
Here and throughout the rest of the paper, $a_s(\mu)\equiv\alpha_s(\mu)/(4\pi)$ is the strong
coupling, its running is
$\beta(a_s)\equiv\dd a_s/\dd\ln\mu^2 = -\beta_0 a_s^2-\beta_1 a_s^3-\cdots$
with $\beta_0=11-2N_f/3$, and the kernels expand as
$\gamma(z;\mu)=\sum_{k\geqslant 1}a_s^k(\mu)\,\gamma^{[k]}(z)$.
Therefore, transformations that satisfy the boundary condition must be
connected to the identity. This simple
condition will play an important role in Sec.~\ref{sec:injectivity}.

\textbf{(iii) Reality:} Because $K(x;\mu)$ is real, its Mellin transform
obeys the Schwarz reflection
$\Kc(\bar z;\mu)=\overline{\Kc(z;\mu)}$. The entries are complex at
generic $z$ but the matrix is real at real $z$. The sum rules act at the
real moments $z=1$ and $z=2$, where $\Kc$ is therefore real.

The above properties combine into a definition that fixes the ``arena'' for
the rest of the paper.

\begin{definition}[Admissible scheme transformation]
\label{def:admissible}
A scheme transformation is a matrix $\Kc(z;\mu)$ on flavor space,
invertible and differentiable in $\mu$ at every Mellin moment and
analytic in $z$ with the Schwarz reflection of property (iii), which
acts on the densities by Eq.~\eqref{eq:change_scheme}, carries the
kernels along by Eq.~\eqref{eq:gauge_transformation}, and approaches
the identity perturbatively, $\Kc=\Id+O(a_s)$, as in
Eq.~\eqref{eq:asymptotic_identity}. We call the transformation
admissible when the transformed kernel obeys the symmetry constraints
of Eq.~\eqref{eq:kernel_symmetries} and the transformed densities obey
the sum rules \eqref{eq:momentum_sum_rule_x} and
\eqref{eq:number_sum_rule_x} whenever the input densities do.
\end{definition}

We describe the analyticity requirement of
Definition~\ref{def:admissible} in more detail, because it fixes the
function space in which the whole classification takes place. The
requirement is the Mellin-space image of support on the unit interval.
If a kernel lives on $0\leqslant x\leqslant1$, say an integrable
function plus a multiple of $\delta(1-x)$, its Mellin transform is
analytic to the right of some vertical line and does not grow
exponentially in the imaginary direction, because $|x^{z-1}|$ depends
only on the real part of $z$. A function with these two properties is
fixed by its values at the positive integers. This is Carlson's theorem
\cite{Boas:1954}, and the reason is simple. Two such functions with the
same integer values differ by a function that vanishes at every
integer, and a function that vanishes at every integer without
vanishing identically behaves like $\sin\pi z$, which grows
exponentially in the imaginary direction. The support forbids exactly
that growth. Two transformations with the same integer moments are
therefore the same transformation.

In particular, a transformation can be prescribed freely at finitely
many moments, as the sum rules do at $z=1$ and $z=2$, but not moment
by moment, since a generic assignment of a matrix to every $z$ has no
$x$-space image. For the same reason, conditions imposed at all
integer moments are exactly as strong as $x$-space statements, which
is what will allow Sec.~\ref{sec:positivity} to characterize pointwise
positivity through moments alone, and statements established on an
open region of the $z$ plane extend to the whole half-plane, a
continuation that the completeness argument of Sec.~\ref{sec:injectivity}
uses in an essential way.

We are now ready to ask precise questions. Which transformations are
admissible in the sense of Definition~\ref{def:admissible}? We impose
the symmetries first and the sum rules second.

\section{The classification}
\label{sec:classification}

\subsection{Enforcing charge conjugation and flavor symmetry}
\label{sec:symmetries}

We first impose that the transformed kernel $\gamma^{(s)}$ satisfy the
symmetry constraints of Sec.~\ref{sec:qcd_symmetries} at every moment.
Since those constraints say that a kernel commutes with the group
$G=\langle C\rangle\times S_{N_f}$, the natural condition on the
transformation is that it commute with $G$ as well,
\begin{equation}
    [\Kc,\,C]=0,
    \qquad
    [\Kc,\,\sigma]=0\quad\forall\,\sigma\in S_{N_f} .
    \label{eq:commutant_condition}
\end{equation}
This condition is clearly sufficient, since both terms of
Eq.~\eqref{eq:gauge_transformation} then commute with $G$ separately
(the $\mu$ derivative does not spoil commutation with constant
matrices). Whether it is also necessary is a separate question, because
one could imagine the two terms violating the symmetry in compensating
ways. This point is taken up in Sec.~\ref{sec:injectivity}, which shows
that no such compensation can occur, so the commutant condition is
necessary as well.

Matrices commuting with a symmetry group are computed by Schur's lemma
from the decomposition of flavor space into irreducible representations
of $G$. The first line below is the two-dimensional subspace spanned by
the gluon density $g$ and the $C$-even quark singlet $\Sigma_+$. Both are
invariant under all of $G$, so the trivial irrep occurs twice and an
arbitrary $2\times2$ mixing between them is allowed. The remaining lines
are sectors that each carry a single irreducible representation, of
dimension one for the $C$-odd singlet and $N_f-1$ for each traceless
non-singlet, so Schur's lemma admits only a scalar multiple of the
identity on each:
\begin{widetext}
\begin{equation}
\begin{array}{llll}
    \{\,g,\;\Sigma_+\,\},
    & \Sigma_\pm \equiv \textstyle\sum_i (q_i\pm\bar q_i),
    & \text{two copies of the trivial irrep}
    & \Rightarrow\ \GL(2)\ \text{block},\\[3pt]
    \Sigma_-,
    & \text{$C$-odd singlet}
    &
    & \Rightarrow\ \GL(1),\\[3pt]
    (q_i+\bar q_i)\ \text{traceless part},
    & \text{$C$-even non-singlet}
    &
    & \Rightarrow\ \GL(1),\\[3pt]
    (q_i-\bar q_i)\ \text{traceless part},
    & \text{$C$-odd non-singlet}
    &
    & \Rightarrow\ \GL(1).
\end{array}
\label{eq:irrep_decomposition}
\end{equation}
\end{widetext}
The commutant is thus
\begin{equation}
    \Kc(z;\mu)\;\in\;\GL(2)\times\GL(1)^3
    \qquad\text{at every Mellin moment,}
    \label{eq:commutant_arena}
\end{equation}
of dimension $7$. Note that $7$ is exactly the number of independent
kernel functions in Eq.~\eqref{eq:kernel_symmetries}. In this sense the
commutant condition is nothing but the symmetry structure of the kernels
transplanted to $\Kc$. 

To be clear, in the parton basis no entry of $\Kc$ is forced
to vanish. The symmetries equate entries there, as in
Eq.~\eqref{eq:kernel_symmetries}, and the dimension counts independent
functions and not surviving entries. The vanishing blocks appear only
in the basis of Eq.~\eqref{eq:irrep_decomposition}, where mixing
between inequivalent representations is forbidden and each irreducible
sector carries a single scalar. (For $N_f=1$ the non-singlet sectors are absent
and the commutant is $\GL(2)\times\GL(1)$, of dimension $5$. At generic
complex $z$ the blocks are complex, with the Schwarz reflection of
Sec.~\ref{sec:mellin}(iii) tying them to real forms on the real axis.)

To summarize, at every Mellin moment a symmetry-preserving scheme
transformation is reduced to four independent pieces, a $2\times2$ block mixing the
quark singlet with the gluon and one scalar for each of the remaining
sectors. In one sentence, the symmetries forbid a scheme from mixing
inequivalent irreps, and only partons that transform identically can be
mixed. The singlet-gluon block exists precisely because $\Sigma_+$ and
$g$ carry the same irrep. Whatever the sum rules do, they must do it
within these blocks.

\subsection{Sum rules as covector conditions}
\label{sec:covectors}

In Mellin space the two sum rules become statements about two special
moments. Let $e_j^\dagger$ be the elementary unit covector that selects
the density of parton $j$, meaning
$e_j^\dagger\,\overline f=\overline f_j$, with the label $j$ running
over the gluon $g$ and the quark flavors $q_i$ and $\bar q_i$. Each sum
rule then comes with its own constant covector on flavor space, a
momentum covector $q_M^\dagger$ that sums over all the partons and one
number covector $q_{N,i}^\dagger$ per flavor, the subscripts standing
for momentum and number:
\begin{equation}
\begin{gathered}
    q_M^\dagger \equiv \sum_j e_j^\dagger = (1,1,\dots,1)
    \quad \text{acting at } z=2,\\
    q_{N,i}^\dagger \equiv e_{q_i}^\dagger - e_{\bar q_i}^\dagger
    \quad \text{acting at } z=1 .
\end{gathered}
    \label{eq:sum_rule_covectors}
\end{equation}
The sum rules read $q_M^\dagger\overline f(2;\mu)=1$ and
$q_{N,i}^\dagger\overline f(1;\mu)=n_{q_i}$, and their scale independence
is equivalent to the covectors being annihilated by the kernel at the
corresponding moment
\begin{equation}
    q_M^\dagger\,\gamma(2;\mu)=0,
    \qquad
    q_{N,i}^\dagger\,\gamma(1;\mu)=0 .
    \label{eq:covector_annihilation}
\end{equation}

We now ask how this behaves under a change of scheme. A short computation
with Eq.~\eqref{eq:gauge_transformation} gives
\begin{equation}
    q^\dagger\,\gamma^{(s)}
    = \left(q^\dagger \Kc\,\gamma
    + \frac{\dd\,(q^\dagger\Kc)}{\dd\ln\mu^2}\right)\Kc^{-1},
    \label{eq:one_line}
\end{equation}
so the sum rule survives in the new scheme precisely when $q^\dagger\Kc$
is again a constant covector annihilated by $\gamma$. At the momentum
moment the space of such covectors is spanned by $q_M^\dagger$ itself.
At the number moment it is spanned by all the $q_{N,i}^\dagger$
together, but the valence numbers of different flavors vary
independently from hadron to hadron, so a sum rule that holds with the
same values for every hadron leaves no room to mix them. In both cases
we need
\begin{equation}
    q^\dagger\,\Kc = \eta\, q^\dagger
    \qquad \text{with } \eta \text{ a $\mu$-independent constant.}
    \label{eq:left_eigenvector}
\end{equation}
That is, $q^\dagger$ must be a left eigenvector of $\Kc$ at the relevant
moment. The eigenvalue is then fixed by the normalization.

Applying
Eq.~\eqref{eq:left_eigenvector} to the densities we find
$q^\dagger\overline f^{(s)} = \eta\, q^\dagger\overline f$, so a
transformation with $\eta\neq1$ rescales the conserved charge. A scheme
in which the proton carries $87\%$ of its own momentum, or $2.3$ valence
up quarks, is simply not a scheme in which the sum rules hold. Hence
\begin{equation}
    \eta = 1 \ \text{ for every sum-rule covector.}
    \label{eq:eta_one}
\end{equation}
The same conclusion follows from the boundary condition
\eqref{eq:asymptotic_identity} alone, since $\eta$ is $\mu$ independent
and $\Kc\to\Id$, so the eigenvalue can only be one.

At order $a_s$ these conditions are well known. The vanishing first moment
of the quark kernel was enforced through the Adler sum rule in
Refs.~\cite{Altarelli:1978,Altarelli:1979}, and the second moment of the
DIS-scheme gluon was fixed by momentum conservation in
Ref.~\cite{Altarelli:1979}. The general order-$a_s$ statement, that a
change of scheme shifts the number sum rule by the first moment of the
$qq$ kernel, was derived recently in Ref.~\cite{Delorme:2025}.

The
condition also has an operator-level counterpart. In minimal subtraction
the number sum rule survives renormalization because the first moment of
the PDF counterterm is trivial,
$\int_0^1\dd\xi\,Z^{\rm pdf}_{qj}(\xi)=\delta_{qj}$, a fact used
recently in Ref.~\cite{Rogers:2026} to constrain the choice of
factorization scale. There the condition falls on the singular
subtraction, here it falls on the finite transformation.
Equations~\eqref{eq:left_eigenvector} and \eqref{eq:eta_one} are the
all-orders version of these statements.

We stress that schemes violating
them are not inconsistent, and several schemes in current use do violate
them \cite{Delorme:2025}. In such schemes the sum-rule integrals simply
become scale-dependent perturbative quantities, and the parton-model
reading of the densities is given up. Our classification delimits the
subgroup of transformations for which that reading is preserved exactly.

A consequence of Eq.~\eqref{eq:left_eigenvector} is worth addressing,
because it separates two roles of the factorization scale. A change of
$\mu_F$ implemented through the evolution itself, meaning the
transformation $\Kc = E_\gamma(\mu_F,\mu)$ that connects the densities
at two points of the DGLAP flow, preserves the sum rules automatically.
Since $q^\dagger\gamma=0$ at its moment at every scale,
$q^\dagger E_\gamma = q^\dagger$ exactly, and the same holds for any
truncation, because each $\gamma^{[k]}$ annihilates the covectors
separately. Scale variation therefore never violates a sum
rule. Whether it is admissible in the sense of
Definition~\ref{def:admissible} is a separate question, decided by the
boundary condition \eqref{eq:asymptotic_identity}. Moving the scale by
a fixed ratio approaches the identity as $\mu$ grows and is admissible,
while evolving to a scale held fixed is not, and
Sec.~\ref{sec:scale_variation_relation} makes the distinction explicit.

A
factorization scale introduced instead as an independent scale of the
operator subtraction is a genuine deformation of the scheme, not a
motion along the flow. Ref.~\cite{Rogers:2026} studies this case at
the level of the operator definitions and finds that the sum rules
require the subtraction scales to be matched. We do not pursue it
here.

\subsection{Combining sum rules and symmetries}
\label{sec:final_cut}

Preserving a sum rule therefore means fixing its covector exactly, and we
know where each covector lives in the decomposition
\eqref{eq:irrep_decomposition}. The momentum covector $q_M^\dagger$ has
components only along $\{g,\Sigma_+\}$, so at $z=2$ it cuts the $\GL(2)$
block down to the stabilizer of one covector in $\GL(2)$, which is the
affine group of the line. The number covectors $q_{N,i}^\dagger$ span the
$C$-odd singlet and the entire $C$-odd non-singlet sector, so at $z=1$
both $C$-odd scalars are forced to one. The $C$-even non-singlet scalar
is untouched by any sum rule at any moment.

At the
real moments the blocks are real
by the Schwarz reflection, and both
$\GL(1,\mathbb{R})\cong\mathbb{R}\setminus\{0\}$ and $\Aff(1,\mathbb{R})$
have two connected components. The boundary condition
\eqref{eq:asymptotic_identity}, together with continuity in $\mu$,
confines $\Kc$ to the identity component of each factor. Writing
$\mathbb{R}_{>0}=\GL(1,\mathbb{R})^{+}$ and
$\Aff(1,\mathbb{R})^{+}\cong\mathbb{R}_{>0}\ltimes\mathbb{R}$ (the
orientation-preserving affine maps $x\mapsto cx+b$ with $c>0$), the
complete classification is the following.

\begin{theorem}[Classification]
\label{thm:classification}
A scheme transformation is admissible if and only if it lies in the
commutant $\GL(2)\times\GL(1)^{3}$ of Eq.~\eqref{eq:commutant_arena} at
every moment, each factor lies in its identity component at real
moments, the singlet-gluon block obeys
$q_M^\dagger\,\Kc(2;\mu)=q_M^\dagger$ at $z=2$, and both $C$-odd
scalars equal one at $z=1$. With each factor read at its own moment,
the conditions combine into
\begin{equation}
    \Kc \;\in\;
    \underbrace{\Aff(1,\mathbb{R})^{+}}_{\text{singlet-gluon, } z=2}
    \;\times\;
    \underbrace{\mathbb{R}_{>0}}_{\text{$C$-even non-singlet}}
    \;\times\;
    \underbrace{\{\Id\}^{2}}_{\text{$C$-odd, } z=1}
    \label{eq:final_group}
\end{equation}
for $N_f>1$, while for $N_f=1$ the two non-singlet factors are absent.
\end{theorem}

\noindent
The sufficiency half is the construction of this section, and the
necessity half is Theorem~\ref{thm:injectivity} below. Each factor of
Eq.~\eqref{eq:final_group} must be read at its proper moment. The $\Aff(1,\mathbb{R})^{+}$
factor is the freedom of the singlet-gluon block at $z=2$. The frozen
$C$-odd factors express the number sum rules at $z=1$. The
$\mathbb{R}_{>0}$ non-singlet factor is untouched by any sum rule and
survives at every moment, as a positive scalar on the real axis and as
one analytic function of $z$ elsewhere, that is, as a convolution kernel
in $x$ space. Away from $z=1,2$ the remaining blocks are restricted only
by the commutant structure of Eq.~\eqref{eq:commutant_arena}.
Moment by moment, the chain of constraints is the following.
\begin{widetext}
\begin{center}
\renewcommand{\arraystretch}{1.5}
\begin{tabular}{l|l|l}
\textbf{Constraints imposed} & \textbf{Group} & \textbf{Dim.}\\
\hline
DGLAP covariance only
& $\GL(2N_f{+}1,\mathbb{C})$ at each $z$
& $(2N_f{+}1)^2$\\
$+$ $C$ $+$ flavor symmetry, every $z$
& $\GL(2)\times\GL(1)^3$
& $7$\\
symmetries $+$ momentum sum rule, $\Kc(2;\mu)$
& $\Aff(1,\mathbb{R})^{+}\times\mathbb{R}_{>0}^{\,3}$
& $5$\\
symmetries $+$ number sum rules, $\Kc(1;\mu)$
& $\GL(2,\mathbb{R})^{+}\times\mathbb{R}_{>0}\times\{\Id\}^{2}$
& $5$\\
all constraints, moment-independent $\Kc$
& $\Aff(1,\mathbb{R})^{+}\times\mathbb{R}_{>0}\times\{\Id\}^{2}$
& $3$\\
\end{tabular}
\end{center}
\end{widetext}
The third and fourth rows are the groups at the momentum and at the
number moment, where only one sum rule is active, and for $N_f=1$ the
two non-singlet scalars are absent and the dimensions become $5$, $3$,
$4$ and $2$.

Equivalently, Eq.~\eqref{eq:final_group} is exactly the group of
moment-independent admissible transformations, which carries three free
functions of $\mu$ (two for $N_f=1$). Admittedly, no practical scheme
is moment independent. In $x$ space such a transformation is a pointwise
reweighting, $K(x;\mu)=A(\mu)\,\delta(1-x)$, with no convolution
structure. In operator language, it assigns the same finite
renormalization to the spin-$z$ operator at every $z$, whereas genuine
scheme changes renormalize each spin independently. The value of this
special case is purely diagnostic. It is the case where the tension between the differential symmetry
constraints and the algebraic sum-rule constraints is strongest, and it
is exactly where the completeness argument of the next section is put to
the test.

The moment-independent commutant has a second use, which turns the
theorem into a recipe. Suppose a transformation lies in the commutant and tends to the identity but
fails one or both sum rules. The failure consists of numbers, the two
$C$-odd scalars of $\Kc(1;\mu)$ and the covector
$q_M^\dagger\Kc(2;\mu)$ in the singlet-gluon block, and it is undone
by a moment-independent element $\Kc_{\mathcal N}(\mu)$ of the
commutant that carries the inverse of each $C$-odd scalar on its
sector and the diagonal matrix
$(n_\Sigma,n_g)=q_M^\dagger\Kc(2;\mu)^{-1}$ on $\{\Sigma_+,g\}$. The
product $\Kc_{\mathcal N}\Kc$ satisfies both covector conditions
exactly and is admissible. The sum rules by themselves are therefore
never an obstruction. A normalization by at most four numbers, which
the sum rules fix, namely the two $C$-odd scalars at $z=1$ and the two
diagonal entries $n_\Sigma$ and $n_g$ at $z=2$, makes any
transformation with the right symmetries and boundary behavior
admissible, and it acts in $x$ space as a pointwise reweighting. At order $a_s$ and for the momentum sum rule
this is the $\delta(1-x)$ counterterm of
Refs.~\cite{Delorme:2025,Delorme:2026}, which leave the number sum
rules modified, and the same device restores them. We exploit this in
Sec.~\ref{sec:tmd}, where it keeps the partonic interpretation of both
the TMDs and the collinear densities defined as their integrals.

The physical reading of Eq.~\eqref{eq:final_group} is that the only
scheme freedoms compatible with all the physics are a reshuffling of
momentum between the quark singlet and the gluon, since only the total
momentum is physical while its split between quarks and gluons is a
convention, and a common positive rescaling
$\nu(\mu)\in\mathbb{R}_{>0}$, with $\nu(\infty)=1$, of the $C$-even
non-singlet combinations, whose normalization can drift but whose sign
can never flip. 
Here the space we consider matters since a moment-independent rescaling
acts pointwise, so it preserves the sign of the distribution at every
$x$, while a general admissible transformation is positive at each real
moment, which protects the sign of every real Mellin moment but not the
pointwise sign in $x$. We return to this point in
Sec.~\ref{sec:positivity}.

The valence number, the $z=1$
moment, is completely fixed but at any other moment the $C$-odd scalars still admit a positive rescaling
at each real moment as allowed by the commutant. Thus, a
moment-dependent scheme
change can reshape a valence distribution in $x$ through a convolution
kernel of unit integral. Only in the moment-independent reading of
Eq.~\eqref{eq:final_group} is the valence sector frozen entirely.

Furthermore, a constant covector annihilated by the transformed kernel at some
moment must annihilate the leading kernel there, because a
transformation with $\Kc=\Id+O(a_s)$ first changes the kernels at
order $a_s^2$, and the leading kernel is therefore the same in every
admissible scheme. Consequently, sum rules can live only at the left null vectors of
$\gamma^{[1]}$, a fact that will be needed in
Sec.~\ref{sec:other_moments}. Whether one holds beyond leading order
is, however, a property of the scheme, since $q^\dagger\Kc^{-1}$ is
annihilated by the transformed kernel only when it is scale
independent, which is exactly the statement that $\Kc$ preserves the
sum rule. Moreover, an admissible transformation can neither create nor destroy
a sum rule at a moment where the leading kernel has no null vector,
and at an accidental leading-order zero it can do either.

It is instructive to note how much bigger the freedom would be without
the symmetries. The set of
invertible matrices fixing $k$ independent covectors is a
block-triangular semidirect product
$\GL(n-k)\ltimes\mathbb{R}^{k(n-k)}$ of dimension $n(n-k)$, with
$n=2N_f+1$. The momentum sum rule alone ($k=1$) would allow the full
group of matrices whose columns sum to one. These are generalized Markov
transition matrices on the momentum fractions, known as the stochastic
group~\cite{Poole1995TheSG}, isomorphic to $\Aff(2N_f,\mathbb{R})$. The
number sum rules alone ($k=N_f$) would allow the group
$\GL(N_f+1,\mathbb{R})\ltimes\mathbb{R}^{N_f(N_f+1)}$. The symmetries
cut these large groups down to the four blocks of
Sec.~\ref{sec:symmetries}, and this is what makes the final answer of
Eq.~\eqref{eq:final_group} so small.

\subsection{Scheme dependence of the momentum fractions}
\label{sec:gluon_fraction}

As a first application, the $\Aff(1,\mathbb{R})^{+}$ factor of
Eq.~\eqref{eq:final_group} acts on the momentum fractions themselves.
Since $\Sigma_+$ and $g$ sit in the same $2\times2$ block, their
fractions have to be treated together. Restricted to the
$\{\Sigma_+,g\}$ plane at $z=2$, an admissible $\Kc$ is a $2\times2$
matrix whose columns sum to one, so in the basis $(\Sigma_+,g)$ it can
be parametrized by its second row,
\begin{equation}
    \Kc\big|_{z=2}
    = \begin{pmatrix} 1-a(\mu) & 1-b(\mu) \\[2pt] a(\mu) & b(\mu) \end{pmatrix},
    \quad a(\infty)=0,\ b(\infty)=1 .
    \label{eq:K_z2}
\end{equation}
Acting with it on the pair of momentum fractions
$\langle x\rangle_j \equiv \overline f_j(2;\mu)$ for
$j=\Sigma_+,g$, and using the
momentum sum rule $\langle x\rangle_{\Sigma_+}+\langle x\rangle_g=1$ to
eliminate one variable, the momentum fractions in the new scheme are
\begin{align}
    \langle x\rangle_g^{(s)}
    &= a(\mu)\,\big(1-\langle x\rangle_g\big) + b(\mu)\,\langle x\rangle_g ,
    \label{eq:gluon_momentum_scheme}\\
    \langle x\rangle_{\Sigma_+}^{(s)}
    &= \big(1-a(\mu)\big)\,\langle x\rangle_{\Sigma_+}
     + \big(1-b(\mu)\big)\,\big(1-\langle x\rangle_{\Sigma_+}\big) .
    \label{eq:singlet_momentum_scheme}
\end{align}
The two formulas are not independent, they sum to one identically,
whatever $a(\mu)$ and $b(\mu)$ are. Every unit of momentum taken from the singlet
reappears in the gluon and vice versa. Either fraction can be given any
value at a given scale while the total stays exactly one. The split of
the hadron's momentum between quarks and gluons is thus pure convention, and
Eqs.~\eqref{eq:gluon_momentum_scheme}--\eqref{eq:singlet_momentum_scheme}
are its exact all-orders parametrization. On the contrary, the valence numbers admit no such formula and they are scheme invariants.

\begin{figure*}[!t]
    \centering
    \includegraphics[width=\textwidth]{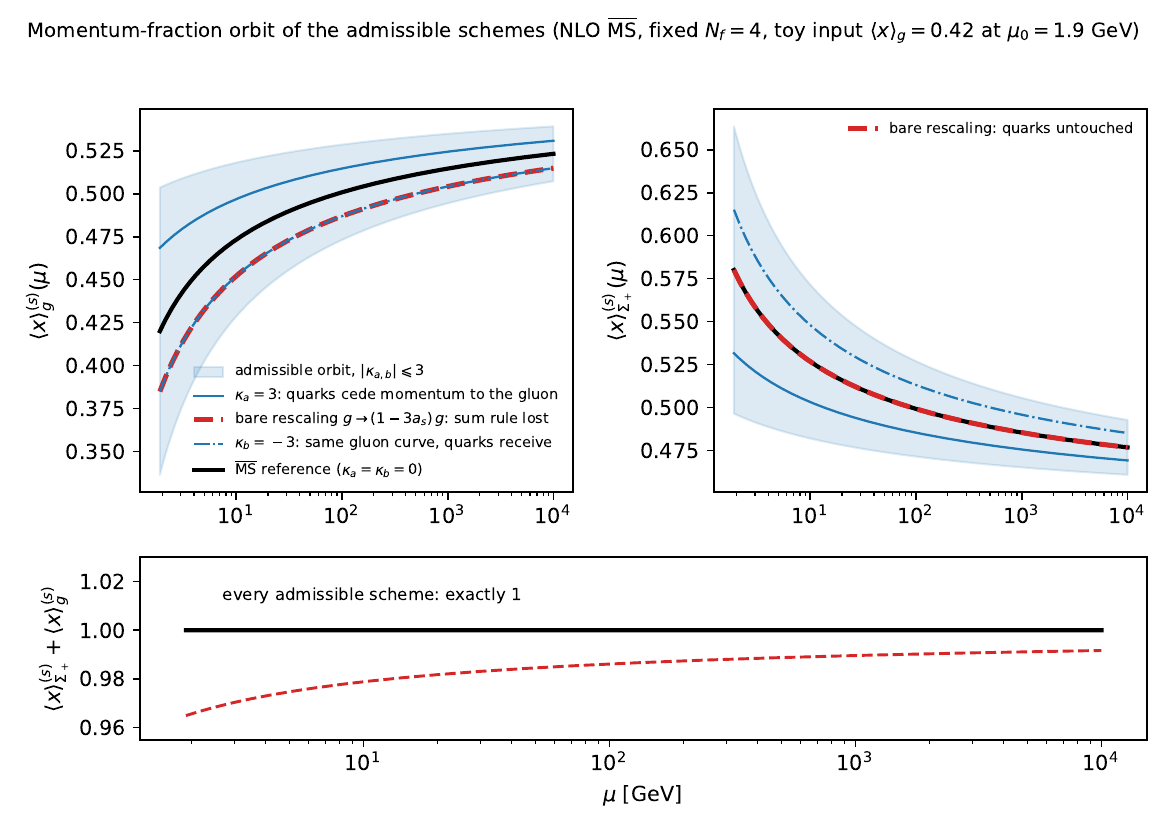}
    \caption{The orbit of the gluon and quark-singlet momentum
    fractions under admissible scheme transformations, computed at NLO
    from a toy input $\langle x\rangle_g=0.42$ at $\mu_0=1.9$~GeV with
    $N_f=4$ held fixed. The upper panels show the reference scheme,
    taken to be $\overline{\rm MS}$, two admissible schemes with
    $a(\mu)=\kappa_a\,a_s(\mu)$ and $b(\mu)=1+\kappa_b\,a_s(\mu)$, and
    the shaded envelope of the family $|\kappa_{a,b}|\leqslant3$, an
    arbitrary illustrative range. The two panels are mirror images of
    each other because the total momentum equals one in every
    admissible scheme. The dashed curve rescales the gluon alone. It
    lies exactly on top of the admissible $\kappa_b=-3$ curve in the
    gluon panel and on top of the reference curve in the singlet panel,
    and the lower panel, which shows the total momentum of
    Eqs.~\eqref{eq:gluon_momentum_scheme}--\eqref{eq:singlet_momentum_scheme},
    separates it from both.}
    \label{fig:orbit}
\end{figure*}

We illustrate this reshuffling of momentum in Fig.~\ref{fig:orbit},
where we evolve a toy input with $\langle x\rangle_g=0.42$ at
$\mu_0=1.9$~GeV at NLO in $\overline{\rm MS}$, which serves as the
reference scheme, with $N_f=4$ held fixed and no heavy-quark
thresholds. The evolution uses the $z=2$ moments of the NLO
$\overline{\rm MS}$ anomalous dimensions as implemented in
Ref.~\cite{Candido:2022}. We then transform the input with the
one-parameter shapes $a(\mu)=\kappa_a\,a_s(\mu)$ and
$b(\mu)=1+\kappa_b\,a_s(\mu)$, where $\kappa_a$ and $\kappa_b$ are
constants of order one. The range $|\kappa_{a,b}|\leqslant3$ used for
the band is an arbitrary choice that only serves as an example of
scheme variation, since nothing in the classification selects a
preferred size for the constants. We will encounter the same shapes later in
Sec.~\ref{sec:scheme_variation} as the minimal scheme variation, with
$a_s$ raised to the working order.

Every
member of the family moves the gluon and quark-singlet fractions in
opposite directions at low scales, so the two upper panels are mirror
images of each other, and all curves collapse onto the reference ones
as $\mu\to\infty$ because of the boundary condition
\eqref{eq:asymptotic_identity}. The lower panel makes
the covector condition visible. The admissible curves keep
$\langle x\rangle_{\Sigma_+}^{(s)}+\langle x\rangle_g^{(s)}$ equal to
one at every scale, as
Eqs.~\eqref{eq:gluon_momentum_scheme}--\eqref{eq:singlet_momentum_scheme}
guarantee.

The dashed curve instead rescales the gluon alone,
$g\to(1-3a_s)\,g$, with no compensating change of the quarks. In the
gluon panel it cannot be told apart from the admissible transformation
with $\kappa_b=-3$, which does the same to the gluon but hands the
momentum to the quarks, and in the singlet panel it cannot be told
apart from the reference curve, which it leaves untouched. Each panel
alone makes the violating scheme look harmless in its own way, and only
the total in the lower panel exposes it.

\section{Completeness of the classification}
\label{sec:injectivity}

\subsection{Necessity of the commutant condition}
\label{sec:necessity}

We now come back to the question left open in Sec.~\ref{sec:symmetries}.
Is the commutant condition \eqref{eq:commutant_condition} necessary as
well as sufficient, or can a transformation outside the commutant produce
a symmetric kernel through a cancellation between the two terms of
Eq.~\eqref{eq:gauge_transformation}? The question is settled by a simple
observation about the boundary condition.

Suppose two admissible transformations $\Kc$ and $\Kc'$ map the same
$\gamma$ to the same $\gamma^{(s)}$. Substituting $\Kc'=\Kc M$ into
Eq.~\eqref{eq:gauge_transformation} and equating, everything involving
$\Kc$ cancels and the ratio must satisfy
\begin{equation}
\begin{gathered}
    \frac{\dd M}{\dd\ln\mu^2} = \left[\gamma,\, M\right],
    \qquad\text{solved by}\\
    M(\mu) = E_\gamma(\mu,\mu_0)\, M(\mu_0)\, E_\gamma(\mu,\mu_0)^{-1},
\end{gathered}
    \label{eq:adjoint_transport}
\end{equation}
with
$E_\gamma(\mu,\mu_0)=\mathcal{T}\exp\big(\int_{\mu_0^2}^{\mu^2}\dd\ln\mu'^2\,
\gamma\big)$ the DGLAP evolution operator, $\mathcal T$ denoting ordering
in $\ln\mu'^2$ and $\mu_0$ a reference scale. Without boundary conditions
any $M(\mu_0)$ is allowed. These form the ``same evolution'' equivalence
classes of transformations. The simplest members are the constant
multiples of the identity, $M = a\,\Id$, the center of
$\GL(2N_f{+}1,\mathbb{C})$. They trivially rescale every density by the same
factor and leave the kernel exactly invariant, as the linearity of the
evolution demands. Note, however, that a constant $a\neq1$ never approaches the
identity, so the boundary condition \eqref{eq:asymptotic_identity} excludes them.

In fact, the boundary condition eliminates all the members order by order in
the coupling. An admissible transformation approaches the identity
perturbatively, as a power series in $a_s$ with coefficients analytic
in $z$, and so does $M$,
\begin{equation}
    M = \Id + \sum_{k\geqslant1} a_s^{\,k}\, m_k(z) .
    \label{eq:M_series}
\end{equation}
Let $m_k$ be the first coefficient that does not vanish. Inserting the
series into Eq.~\eqref{eq:adjoint_transport}, the left-hand side starts
at order $a_s^{k+1}$, because
$\dd a_s^{k}/\dd\ln\mu^2=k\,a_s^{k-1}\beta(a_s)=-k\beta_0a_s^{k+1}+O(a_s^{k+2})$,
and so does the right-hand side, where the leading kernel meets $m_k$.
At that order
\begin{equation}
    \big[\gamma^{[1]},\,m_k\big] + k\,\beta_0\,m_k = 0 .
    \label{eq:injectivity_operator}
\end{equation}
The map $m\mapsto[\gamma^{[1]},m]+k\beta_0\,m$ is invertible unless
$-k\beta_0$ is an eigenvalue of the commutator with $\gamma^{[1]}$. On
the scalar sectors of Eq.~\eqref{eq:irrep_decomposition} the commutator
vanishes and the map multiplies by $k\beta_0$. On the singlet-gluon
block the commutator has eigenvalues $0$, twice, and
$\pm\beta_0\Delta(z)$, where $\beta_0\Delta(z)$ is the difference of
the two eigenvalues of $\gamma^{[1]}(z)$, so the map has eigenvalues
$k\beta_0$, twice, and $\beta_0\big(k\pm\Delta(z)\big)$, and it
degenerates only where $\Delta(z)=k$, at isolated moments. Elsewhere
$m_k=0$, and at those moments too by continuity in $z$. No first
nonvanishing coefficient exists, and $M=\Id$. For $N_f=4$ the moments
with $\Delta(z)=1$ are $z\simeq1.80$ and $z\simeq3.85$, and those with
$\Delta(z)=2$ are $z\simeq1.49$ and $z\simeq12.8$. They reappear in
Part~\ref{part:two} as the moments where the series solution of the
evolution becomes degenerate. In one sentence, the running of the
coupling and the leading kernel act on the mismatch at the same order,
and they can cancel only at isolated moments. The formal statement is
the following.

\begin{theorem}[Injectivity]
\label{thm:injectivity}
Let two scheme transformations in the sense of
Definition~\ref{def:admissible} map the same kernel $\gamma$ to the
same transformed kernel. Then they coincide at every moment, so the map
\begin{equation}
    \Kc \mapsto \left(\gamma\Rightarrow\gamma^{(s)}\right)
    \qquad\text{is injective.}
    \label{eq:injectivity}
\end{equation}
\end{theorem}

\noindent
A change of scheme is completely determined by what it does to the
kernel. There is no residual gauge redundancy in
Eq.~\eqref{eq:gauge_transformation}.

Necessity of the commutant now follows from a short symmetry argument.
Conjugate Eq.~\eqref{eq:gauge_transformation} by a symmetry $h\in G$ (we
write $h$ instead of $g$ to avoid confusion with the gluon label). Since
$h$ is constant and both $\gamma$ and $\gamma^{(s)}$ commute with it, the
conjugated equation says that $\Kc_h\equiv h\Kc h^{-1}$ transforms the
same $\gamma$ into the same $\gamma^{(s)}$, and it obeys the same
boundary condition. Injectivity then forces $\Kc_h=\Kc$ for every $h$,
which is exactly Eq.~\eqref{eq:commutant_condition}. The classification
\eqref{eq:final_group} is therefore an if and only if statement. No
admissible transformation exists outside it.

Injectivity also resolves a puzzle of counting: imposing the symmetry
constraints on $\gamma^{(s)}$ directly yields first-order differential
conditions on $\Kc$, and for moment-independent transformations a naive
balance of equations against unknowns is strictly negative, suggesting no
solutions at all. Yet the moment-independent members of
\eqref{eq:final_group} manifestly exist. The resolution is that counting
assumes the constraints to be independent, which holds only at generic
points. On the commutant both terms of the transformation law commute
with $G$ separately, so the differential constraints vanish identically
and cut nothing. The solutions of an over-determined system live exactly
where the constraint map degenerates, and here that region admits a
closed-form description. The moment-independent solutions are precisely
$\Aff(1,\mathbb{R})^{+}\times\mathbb{R}_{>0}$, three free functions of
$\mu$, and nothing else.

We stress that the argument is perturbative. It uses the expansion
\eqref{eq:M_series}, which is what Definition~\ref{def:admissible}
means by approaching the identity perturbatively, and the requirement
is not idle. At leading order the general solution of
Eq.~\eqref{eq:adjoint_transport} contains, in the eigenbasis of
$\gamma^{[1]}$, an off-diagonal piece proportional to
$a_s^{\Delta(z)}$, which vanishes as $\mu\to\infty$ and is
nevertheless not a power series in $a_s$ unless $\Delta(z)$ is an
integer, which is exactly where Eq.~\eqref{eq:injectivity_operator}
degenerates. A transformation approaching $\Id$ through such a power,
or more slowly than $a_s$, or containing genuinely non-perturbative
pieces such as powers of $\Lambda^2/\mu^2$, falls outside the reach of
the theorem. The argument also relies on the analyticity of $\Kc$ in
$z$, required in any case for its $x$-space image to exist with the
growth conditions of Sec.~\ref{sec:mellin}. Within perturbation
theory, where any scheme transformation is a series in $a_s$, the
classification is complete.

\subsection{Other Mellin moments}
\label{sec:other_moments}

The sum rules act at $z=1$ and $z=2$ only. The fact that the story ends
there is itself an exact statement with a physical origin. A sum rule at
moment $z$ would require a constant covector annihilated by
$\gamma(z;\mu)$ at every scale, that is, at every order in $a_s$
simultaneously. The $z$-th moment of a parton density is the matrix
element of a local spin-$z$ operator, and a $\mu$-independent moment
corresponds to a conserved current, whose charges form a Lorentz
tensor of rank $z-1$. By the Coleman-Mandula theorem
\cite{Coleman:1967}, the symmetry algebra of an interacting S-matrix
consists of the Poincar\'e generators and Lorentz-scalar internal
charges, and this leaves room for exactly two ranks. A rank-zero
charge is an internal scalar, and QCD has them in the flavor number
currents at $z=1$. A rank-one charge can only be the momentum $P_\mu$
itself, which selects the energy-momentum tensor at $z=2$. From spin
three on the charge would be a tensor of rank two or more, which the
theorem excludes, so at every other moment all eigenvalues of $\gamma$
are genuine anomalous dimensions and the moments run. Moments where
the leading kernel happens to have a vanishing eigenvalue are
different. There the moment moves at the next order in
$\overline{\rm MS}$, and the first moment of the $C$-even non-singlet
kernel is such a case, but by the argument at the end of
Sec.~\ref{sec:final_cut} an admissible scheme can be chosen in which
it is conserved to all orders, in the same way that the Adler-Bardeen
scheme makes the polarized quark spin scale independent
\cite{Ball:1995}. Such a conservation law is a property of the scheme
and not of QCD, and this is the sense in which no scheme
transformation can manufacture a conservation law where QCD has none.

The absence of further sum rules does not make the other moments
arbitrary. At every moment the
transformation must still be DGLAP
covariant, that is, invertible, differentiable in $\mu$, subject to the
boundary condition and to the transformation law
\eqref{eq:gauge_transformation}, and it must live in the commutant
\eqref{eq:commutant_arena} since the symmetries act at every $z$.
Moreover $\Kc(z;\mu)$ must be analytic in $z$ for its $x$-space image to
exist (Sec.~\ref{sec:mellin}), so the moments are tied together and are
not a collection of independent matrices. At generic $z$, conservation is replaced by
parallel transport. The $\mu$-dependent covectors
$q^\dagger(\mu)=q^\dagger(\mu_0)\,E_\gamma(\mu,\mu_0)^{-1}$ have constant
pairing with $\overline f$, which expresses the invertibility of the
evolution, not a conservation law. Pointwise in $z$, the evolution
is a connection in the single variable $\mu$ and can be gauged away
entirely. Part~\ref{part:two} turns precisely this freedom into a
solution method.

\section{Positivity}
\label{sec:positivity}

Whether the densities of a given scheme, and of $\overline{\rm MS}$ in
particular, are pointwise nonnegative in $x$ is a question under active
discussion \cite{Candido:2020,Collins:2021,Candido:2023}, and we do not
enter it here. However, we note that nothing in the classification protects the sign of the densities. Therefore, we ask instead how positivity behaves under the
transformations we have classified, and the short answer is that it is
fragile. An admissible scheme change can create a negative density or
repair one, and only a special subset of transformations carries
positive schemes into positive schemes. We identify this subset in full
for the moment-independent case and as a moment condition in the more interesting and realistic general case.

The fragility is visible already at the level of the momentum sum rule.
We saw in Sec.~\ref{sec:gluon_fraction} that the affine freedom of
Eq.~\eqref{eq:final_group} can assign the gluon any momentum fraction,
including a negative one, and a density with a negative real moment
must itself be negative on a set of nonzero measure in $x$. A scheme that gives the gluon negative momentum
therefore cannot have a nonnegative gluon density, whatever target it
is applied to. Momentum fractions inside $[0,1]$ are thus a necessary
condition for positivity, and the content of this section is how far
that necessary condition is from a sufficient one.

For the transformations of the classification the characterization is
a set of seven conditions, one for each independent kernel
that survives the symmetries. Preservation here means preservation for
every nonnegative input, the notion appropriate for a scheme, which is
defined independently of any particular target, and it is this
universality that makes the conditions necessary as well as
sufficient.

\begin{theorem}[Positivity]
\label{thm:positivity}
Let an admissible transformation have $x$-space sector kernels
$K_{\Sigma\Sigma}$, $K_{\Sigma g}$, $K_{g\Sigma}$ and $K_{gg}$ for the
singlet-gluon block and $K^{+}_{\rm ns}$, $K^{-}_{\rm ns}$ and
$K_{\Sigma_-}$ for the non-singlet and $C$-odd singlet scalars, each a
function plus a multiple of $\delta(1-x)$. The transformation maps
every collection of nonnegative densities into nonnegative densities
exactly when
\begin{equation}
\begin{gathered}
    K_{\Sigma g}\geqslant0,
    \qquad
    K_{g\Sigma}\geqslant0,
    \qquad
    K_{gg}\geqslant0,\\
    K_{\Sigma\Sigma}-K^{+}_{\rm ns}\geqslant
    \big|K_{\Sigma_-}-K^{-}_{\rm ns}\big| ,
\end{gathered}
    \label{eq:positivity_conditions}
\end{equation}
together with
\begin{equation}
    N_f\big(K^{+}_{\rm ns}\pm K^{-}_{\rm ns}\big)
    + \big(K_{\Sigma\Sigma}-K^{+}_{\rm ns}\big)
    \pm \big(K_{\Sigma_-}-K^{-}_{\rm ns}\big)
    \geqslant 0 ,
    \label{eq:positivity_conditions_diag}
\end{equation}
at every $x$, where the absolute value stands for two of the seven
conditions.
\end{theorem}

\noindent
The proof has two steps. An arbitrary matrix kernel $K_{ij}(x)$, with
$i,j$ running over quarks, antiquarks and the gluon, maps every set of
nonnegative densities into nonnegative densities exactly when every
entry is nonnegative. A convolution of nonnegative functions is
nonnegative, which gives one direction. For the other direction,
preservation is required for every nonnegative input, and a density
carried by a single parton species and concentrated at one momentum
fraction isolates the kernel entries one at a time. A negative entry
anywhere produces a negative density for some target. This
characterization holds for any transformation, whether or not it
respects the symmetries.

For a transformation in the commutant the entries are built from the
seven sector kernels, with the block acting as
$\Sigma_+\mapsto K_{\Sigma\Sigma}\otimes\Sigma_+ + K_{\Sigma g}\otimes g$
and $g\mapsto K_{g\Sigma}\otimes\Sigma_+ + K_{gg}\otimes g$, and the
scalars multiplying their sectors as Mellin convolutions. Decomposing
$q_i^{\pm}=q_i\pm\bar q_i$ into the singlets $\Sigma_\pm$ and the
traceless non-singlet parts, the quark and gluon rows of the matrix
read
\begin{widetext}
\begin{align}
    K_{q_iq_j} &= \frac{\delta_{ij}}{2}\big(K^{+}_{\rm ns}+K^{-}_{\rm ns}\big)
    + \frac{1}{2N_f}\Big[\big(K_{\Sigma\Sigma}-K^{+}_{\rm ns}\big)
    + \big(K_{\Sigma_-}-K^{-}_{\rm ns}\big)\Big] ,
    \nonumber\\
    K_{q_i\bar q_j} &= \frac{\delta_{ij}}{2}\big(K^{+}_{\rm ns}-K^{-}_{\rm ns}\big)
    + \frac{1}{2N_f}\Big[\big(K_{\Sigma\Sigma}-K^{+}_{\rm ns}\big)
    - \big(K_{\Sigma_-}-K^{-}_{\rm ns}\big)\Big] ,
    \nonumber\\
    K_{q_i g} &= \frac{1}{2N_f}\,K_{\Sigma g} ,
    \qquad
    K_{g q_i} = K_{g \bar q_i} = K_{g\Sigma} ,
    \label{eq:parton_basis_action}
\end{align}
\end{widetext}
while the gluon-gluon entry coincides with the sector kernel $K_{gg}$
and the $\bar q_i$ rows follow from charge conjugation,
$K_{\bar q_i\bar q_j}=K_{q_iq_j}$ and
$K_{\bar q_iq_j}=K_{q_i\bar q_j}$. Charge conjugation and flavor
symmetry leave exactly seven distinct entries. The three gluon entries
give the first three conditions of
Eq.~\eqref{eq:positivity_conditions}. The two off-diagonal quark
entries, $i\neq j$, require
$(K_{\Sigma\Sigma}-K^{+}_{\rm ns})\pm(K_{\Sigma_-}-K^{-}_{\rm ns})\geqslant0$,
which is the absolute-value condition. The two diagonal entries
$K_{q_iq_i}$ and $K_{q_i\bar q_i}$, multiplied by $2N_f$, are the two
signs of Eq.~\eqref{eq:positivity_conditions_diag}. For $N_f=1$ the
non-singlet kernels are absent and the conditions reduce to
$K_{\Sigma\Sigma}\geqslant|K_{\Sigma_-}|$ together with the
nonnegativity of $K_{\Sigma g}$, $K_{g\Sigma}$ and $K_{gg}$.

The gluon must be fed and preserved with nonnegative
weight, the singlet-singlet kernel must dominate the non-singlets
pointwise, and the two remaining combinations keep the diagonal flavor
entries nonnegative. For a fixed target the
conditions remain sufficient but are no longer necessary, since a
negative kernel entry can be outweighed by the contributions of the
other partons of that particular target, and it is exposed only by
inputs that switch those contributions off. A transformation that
feeds the singlet a slightly negative gluon kernel leaves a
proton-like input positive because the quark contribution dominates,
and fails only on a nearly pure-gluon target.

We can compose the transformations but not invert them, and that is why
they form a semigroup inside the group \eqref{eq:final_group}, not a
subgroup. Positivity is therefore the opposite of the sum rules
in this respect. The sum rules are preserved by the whole group and are
invariants of the scheme orbit, while positivity marks a region of
scheme space with one-way doors.

In Mellin space the requirement is much stronger than positivity of the
moments. Each of the seven conditions requires one combination of
kernels to be nonnegative on $[0,1]$, so it is enough to characterize
a single nonnegative kernel $K(x)$ through its Mellin transform of
Sec.~\ref{sec:mellin} restricted to the integer moments,
\begin{equation}
    \Kc(n;\mu) = \int_0^1\dd x\,x^{n-1}K(x) ,
    \qquad z = n = 1,2,3,\dots ,
    \label{eq:integer_moments}
\end{equation}
where $\Kc$ stands for the combination considered. The identity components of
Sec.~\ref{sec:final_cut} make the sector kernels positive at each real
moment, and one might hope that positive moments already keep a kernel
nonnegative in $x$. It does not. A kernel with a narrow negative dip
barely changes any of its moments, because the smooth weights $x^{n-1}$
average over the dip, so all the $\Kc(n;\mu)$ can stay positive while
the kernel, and with it a transformed density, turns negative on a
whole interval.

The exact requirement is the complete monotonicity of
the sequence $\Kc(n;\mu)$, which by Hausdorff's moment theorem
\cite{Hausdorff1921} characterizes nonnegative kernels on $[0,1]$ and
demands
\begin{equation}
    (-1)^k\big(\Delta^k\Kc\big)(n;\mu)\geqslant0
    \qquad\text{for every } n\geqslant1 \text{ and } k\geqslant0 ,
    \label{eq:complete_monotonicity}
\end{equation}
where $\Delta^k$ denotes the $k$-fold iteration of the finite
difference and not a power,
\begin{align}
    \big(\Delta^0\Kc\big)(n;\mu) &= \Kc(n;\mu) ,
    \nonumber\\
    \big(\Delta^1\Kc\big)(n;\mu) &= \Kc(n+1;\mu)-\Kc(n;\mu) ,
    \nonumber\\
    \big(\Delta^2\Kc\big)(n;\mu) &= \Kc(n+2;\mu)-2\,\Kc(n+1;\mu)+\Kc(n;\mu) ,
    \nonumber\\
    &\;\;\vdots
    \nonumber\\
    \big(\Delta^k\Kc\big)(n;\mu) &= \sum_{m=0}^{k}(-1)^{k-m}\binom{k}{m}\,\Kc(n+m;\mu) .
    \label{eq:iterated_differences}
\end{align} For a
nonnegative kernel the condition is transparent, because each
difference inserts a factor $x-1$ under the integral and
\begin{equation}
    (-1)^k\big(\Delta^k\Kc\big)(n;\mu)
    =\int_0^1\dd x\;x^{n-1}(1-x)^k\,K(x)\geqslant0
    \label{eq:cm_integral}
\end{equation}
term by term. Moment positivity is only the first layer $k=0$. The
conditions apply to resummed kernels, since fixed-order kernels contain
endpoint distributions with no definite sign.

As a simple example, we consider the moment-independent
transformations. There the surviving freedom can be written down
completely, and it is remarkably small. Of the three functions
$a(\mu)$, $b(\mu)$ and $\nu(\mu)$ of Eq.~\eqref{eq:final_group},
positivity eliminates all but one,
\begin{equation}
    g \mapsto b(\mu)\,g,
    \qquad
    q_i \mapsto q_i + \frac{1-b(\mu)}{2N_f}\,g,
    \qquad
    0 < b(\mu) \leqslant 1 ,
    \label{eq:positivity_semigroup}
\end{equation}
with $b(\infty)=1$ and compositions multiplying the parameters. The
gluon can only give momentum away, the quarks and antiquarks of all
flavors receive it in equal shares, and the non-singlet rescaling is
frozen at $\nu(\mu)=1$.

To confirm this, we insert the kernels
$K_{\Sigma\Sigma}=\big(1-a(\mu)\big)\,\delta(1-x)$,
$K_{\Sigma g}=\big(1-b(\mu)\big)\,\delta(1-x)$,
$K_{g\Sigma}=a(\mu)\,\delta(1-x)$, $K_{gg}=b(\mu)\,\delta(1-x)$,
$K^{+}_{\rm ns}=\nu(\mu)\,\delta(1-x)$ and
$K_{\Sigma_-}=K^{-}_{\rm ns}=\delta(1-x)$ into the conditions
\eqref{eq:positivity_conditions} and
\eqref{eq:positivity_conditions_diag}, which reduce to
\begin{equation}
\begin{gathered}
    a(\mu)\geqslant0,
    \qquad
    b(\mu)\leqslant1,\\
    \nu(\mu)+a(\mu)\leqslant1,
    \qquad
    (\nu(\mu)-1)(N_f-1)\geqslant a(\mu).
\end{gathered}
    \label{eq:positivity_inequalities}
\end{equation}
The condition $K_{gg}\geqslant0$ also returns $b(\mu)\geqslant0$,
which adds nothing new, since the first condition gives
$a(\mu)\geqslant0$ and the identity component keeps the determinant
$b-a$ of Eq.~\eqref{eq:K_z2} positive.
The last two conditions pin $\nu(\mu)=1$ and $a(\mu)=0$ (for $N_f=1$
the non-singlet kernels are absent and the reduced conditions force
$a(\mu)=0$ directly). The inverse would need $b>1$ and a negative quark-gluon
entry, so even this one-parameter family moves only one way.
Keeping only the second moments inside $[0,1]$, by contrast, would
allow the whole stochastic region $0\leqslant a,b\leqslant1$ of
Sec.~\ref{sec:gluon_fraction}, so pointwise positivity in $x$ is
strictly stronger than positivity of the momentum fractions.

The valence sector shows most clearly what such transformations do. The
number sum rule fixes the integral of the $C$-odd kernel to one, so a
nonnegative kernel there is a probability density in the
momentum-fraction ratio, and the transformation acts on a valence
distribution the way one round of parton branching does. Each quark
keeps its identity and hands away a random share of its momentum. Every
real moment beyond $z=1$ of such a kernel lies below one, so a
positivity-preserving scheme change can only move valence quarks toward
smaller $x$, never back.

The evolution itself gives the semigroup a physical reading. Every
admissible transformation approaches the identity at large $\mu$, so
all schemes of the classification share the same densities as
$\mu\to\infty$, and whether the densities are positive in that limit is
a scheme-independent question. At finite scales the schemes differ, and
the bridge between the two regimes is the evolution operator, which
preserves the sum rules and, for a fixed scale ratio, is itself a
member of the classification (Sec.~\ref{sec:covectors}). At
leading order the evolution operator belongs to the positivity
semigroup in every scheme at once. The transformation law
\eqref{eq:gauge_transformation} first modifies the kernels at order
$a_s^2$, so the leading-order kernel is common to the whole
classification, its splitting functions are nonnegative below $x=1$,
and the endpoint distributions exponentiate into a positive overall
factor. Upward evolution at this order is exactly the parton branching
described below Eq.~\eqref{eq:positivity_semigroup}.

Belonging to the semigroup is a conditional statement.
It guarantees nonnegative outputs only for nonnegative inputs, so
leading-order evolution preserves the positivity of the input scale
when it is there and cannot create it when it is not. Downward
evolution instead leaves the semigroup at every order and in every
scheme. In the valence sector the inverse operator has unit first
moment, by the number sum rule, while its second moment exceeds one,
because the moments that shrink on the way up must grow on the way
down, and no nonnegative kernel on $[0,1]$ has increasing real Mellin
moments. The familiar statement that backward evolution can generate
negative densities is this one-way door. The first place where
positivity can distinguish one scheme from another is next-to-leading
order, where the kernels become scheme dependent and their entrywise
positivity is lost, and that is exactly where the discussion of
Refs.~\cite{Candido:2020,Collins:2021,Candido:2023} takes place.

A complete description of the $z$-dependent semigroup, one completely
monotone moment sequence per sector subject to the conditions of
Sec.~\ref{sec:final_cut} at $z=1$ and $z=2$, remains open, and it would
connect the classification directly to the schemes designed to make the
densities positive \cite{Candido:2020,Collins:2021,Candido:2023}.

The same argument covers the polarized densities, where the physical
bounds tie the unpolarized, helicity and transversity distributions
together, and Appendix~\ref{sec:soffer} states the result.

\section{Collinear densities from TMD integrals}
\label{sec:tmd}

It was shown recently how to relate a collinear density to its TMD
counterpart by integrating the
latter over transverse momentum, with a cutoff or a weight that makes
the integral converge
\cite{Gonzalez-Hernandez:2023,Aslan:2024,delRio:2024,Rainaldi:2025efq}.
In the language of this paper that relation is a change of scheme. It
preserves the DGLAP form of the evolution, so it is a transformation of
the kind introduced in Sec.~\ref{sec:mellin}, but whether it is
admissible is a separate question. It is a natural scheme to consider,
because in the parton model the integral of the TMD counts the partons
that carry a given momentum fraction, and it is the definition that
keeps TMDs and collinear densities tied together in phenomenology. The
question is whether it preserves the sum rules.

At small
transverse distance the TMD has an operator product expansion in terms
of, say, the $\overline{\rm MS}$ densities, and integrating it with any cutoff
or weight gives back a convolution of those same densities, up to power
corrections in the cutoff. The integrated density is therefore
\begin{equation}
\begin{gathered}
    \overline f^{\,c}(z;\mu) = \Kc_c^{(\mathrm{unrm})}(z;\mu)\,\overline f(z;\mu),\\[3pt]
    \Kc_c^{(\mathrm{unrm})}(z;\mu) = \Id + \sum_{n\geqslant1} a_s^{\,n}(\mu)\, C^{[n]}_{\Delta}(z),
\end{gathered}
    \label{eq:tmd_scheme}
\end{equation}
with kernels that come from the coefficient functions of the expansion
and from the weight. We call this the unnormalized cutoff scheme,
because the integral is taken as it comes, with no factor in front.
The relation holds at every order \cite{Rainaldi:2025efq,delRio:2024}.
The kernels themselves
are known as far as the coefficient functions have been computed,
through three loops for the zeroth transverse-momentum moment
\cite{delRio:2024}. Once the cutoff and the rapidity scale are tied to
$\mu$, the transformation depends on $\mu$ only through $a_s$, so it
approaches the identity as the coupling vanishes, and since the
coefficient functions respect charge conjugation and flavor symmetry
it lies in the
commutant. Everything is in place for Theorem~\ref{thm:classification},
which says that the integral relation preserves the sum rules exactly
when
\begin{equation}
    q_{N,i}^\dagger\, C^{[n]}_{\Delta}(1) = 0,
    \qquad
    q_{M}^\dagger\, C^{[n]}_{\Delta}(2) = 0
    \qquad\text{for every } n .
    \label{eq:tmd_conditions}
\end{equation}

However, nothing forces these to hold, and in general they do not.
What the covector conditions return instead are the numbers of
Sec.~\ref{sec:final_cut}. At $z=1$ the kernel acts on the $C$-odd
sectors by scalars, and for the cutoff scheme at one loop the kernel
is diagonal in flavor, so the two scalars coincide and we write
\begin{equation}
\begin{gathered}
    q_{N,i}^\dagger\,\Kc_c^{(\mathrm{unrm})}(1;\mu) = \eta^{(1)}_q(\mu)\, q_{N,i}^\dagger ,\\[3pt]
    \eta^{(1)}_q(\mu) = 1 + \sum_{n\geqslant1} a_s^{\,n}(\mu)\, c_n ,
\end{gathered}
    \label{eq:tmd_eta}
\end{equation}
with $c_n$ the first moment of the $C$-odd quark kernel at order $n$,
a number fixed by the constant part of the coefficient function and by
the weight. Should the two $C$-odd scalars differ at some order, the
$C$-odd singlet simply carries its own number. The integrated
densities count $\eta^{(1)}_q(\mu)\,n_{q_i}$ valence quarks, a number
that runs with the scale and reaches the integer only as
$\mu\to\infty$. This is exactly the situation of
Sec.~\ref{sec:covectors}, where a sum rule turns into a
scale-dependent perturbative quantity. At $z=2$ the momentum covector
meets the two-by-two singlet-gluon block, and we define
$\eta^{(2)}_q$ and $\eta^{(2)}_g$ as the two numbers with
\begin{equation}
    \Big(\frac{1}{\eta^{(2)}_q(\mu)},\,\frac{1}{\eta^{(2)}_g(\mu)}\Big)\,
    \Kc_c^{(\mathrm{unrm})}(2;\mu)\Big|_{\{\Sigma_+,g\}} = q_M^\dagger ,
    \label{eq:tmd_eta2}
\end{equation}
two linear equations in two unknowns, with a unique solution at every
order because the block is invertible. In words, $1/\eta^{(2)}_q$ and
$1/\eta^{(2)}_g$ are the factors by which the quark singlet and the
gluon have to be rescaled for the momentum carried by all partons to
add up to one again. All three numbers depend on $\mu$ through $a_s$
and equal one at $a_s=0$.

The normalized cutoff scheme is then a one-line definition,
\begin{equation}
    \Kc_c(z;\mu) = \Kc_{\mathcal N}(\mu)\,\Kc_c^{(\mathrm{unrm})}(z;\mu),
    \label{eq:tmd_normalized}
\end{equation}
with $\Kc_{\mathcal N}$ the moment-independent element of the
commutant built from the three numbers. Sector by sector it is
$1/\eta^{(2)}_q$ on $\Sigma_+$, $1/\eta^{(2)}_g$ on $g$, and
$1/\eta^{(1)}_q$ on both $C$-odd sectors. The sum rules say nothing
about the $C$-even non-singlet scalar, and we set it to
$1/\eta^{(2)}_q$ so that every $C$-even quark combination is treated
like the singlet. In the parton basis this reads
\begin{equation}
\begin{gathered}
    \Kc_{\mathcal N}:\quad
    q_i \mapsto \tfrac12\Big(\frac{1}{\eta^{(2)}_q}+\frac{1}{\eta^{(1)}_q}\Big)\,q_i
              + \tfrac12\Big(\frac{1}{\eta^{(2)}_q}-\frac{1}{\eta^{(1)}_q}\Big)\,\bar q_i ,\\[3pt]
    g \mapsto \frac{1}{\eta^{(2)}_g}\, g ,
\end{gathered}
    \label{eq:tmd_fix_parton}
\end{equation}
a $z$-independent linear map. Multiplying Eq.~\eqref{eq:tmd_eta} by
$1/\eta^{(1)}_q$, and reading Eq.~\eqref{eq:tmd_eta2} as
$q_M^\dagger\Kc_{\mathcal N}\Kc_c^{(\mathrm{unrm})}(2)=q_M^\dagger$,
shows that $\Kc_c$ satisfies both covector conditions exactly. It is
in the commutant and tends to the identity with the coupling, so it is
admissible by Theorem~\ref{thm:classification}. This is the
normalization of Sec.~\ref{sec:final_cut} applied to the cutoff
scheme. The map is a plain rescaling of each species, all quarks and
antiquarks by one factor and the gluon by another, exactly when
$\eta^{(1)}_q=\eta^{(2)}_q$, that is, when the number deficit and the
momentum deficit of a quark coincide. When they do not,
$\Kc_{\mathcal N}$ mixes $q_i$ with $\bar q_i$ by the difference.
Written for the densities themselves, with the cutoff at $\mu$ and the
rapidity scale $\zeta=\mu^2$, the normalized scheme is
\begin{equation}
\begin{aligned}
    (q_i+\bar q_i)(x;\mu) &= \frac{1}{\eta^{(2)}_q}
      \int_{|\boldsymbol k_T|<\mu}\dd^2\boldsymbol k_T\,
      (q_i+\bar q_i)(x,\boldsymbol k_T;\mu,\mu^2),\\[3pt]
    (q_i-\bar q_i)(x;\mu) &= \frac{1}{\eta^{(1)}_q}
      \int_{|\boldsymbol k_T|<\mu}\dd^2\boldsymbol k_T\,
      (q_i-\bar q_i)(x,\boldsymbol k_T;\mu,\mu^2),\\[3pt]
    g(x;\mu) &= \frac{1}{\eta^{(2)}_g}
      \int_{|\boldsymbol k_T|<\mu}\dd^2\boldsymbol k_T\,
      g(x,\boldsymbol k_T;\mu,\mu^2),
\end{aligned}
    \label{eq:tmd_admissible_all}
\end{equation}
valid at every order. The $C$-even quark combinations and the gluon
carry the momentum numbers and the $C$-odd combinations carry the
number one, and each is a plain integral of the TMD up to its
normalization.

To see the size of the effect, take the same cutoff definition at one
loop, where the kernel is the constant part of the coefficient
functions
\cite{Echevarria:2016,Gonzalez-Hernandez:2023,Aslan:2024},
\begin{equation}
\begin{gathered}
    C^{[1]}_{\Delta,qq} = 2C_F\Big[(1-x) - \frac{\pi^2}{12}\,\delta(1-x)\Big],\\[3pt]
    C^{[1]}_{\Delta,qg} = 4T_F\,x(1-x),
    \qquad
    C^{[1]}_{\Delta,gq} = 2C_F\,x,\\[3pt]
    C^{[1]}_{\Delta,gg} = -\frac{\pi^2}{6}\,C_A\,\delta(1-x),
\end{gathered}
\label{eq:tmd_kernels}
\end{equation}
for each quark or antiquark flavor. At this order the three numbers
are one plus $a_s$ times a moment of the kernels, up to terms of order
$a_s^2$: the first moment of the quark kernel for $\eta^{(1)}_q$,
and the column sums of the block at $z=2$ for $\eta^{(2)}_q$ and
$\eta^{(2)}_g$. The first moment of the quark kernel is
$c_1=C_F(1-\pi^2/6)$, so
\begin{equation}
    \eta^{(1)}_q(\mu) = 1 + a_s(\mu)\,C_F\Big(1-\frac{\pi^2}{6}\Big) + O(a_s^2),
    \label{eq:tmd_eta_q1}
\end{equation}
and the integrated valence numbers fall short of the integers by
$2.4\%$ at $\alpha_s=0.35$ and by $0.8\%$ at $\alpha_s=0.118$. The
quark column at $z=2$ gives the same number, because the regular
parts of the two kernels fed by a quark add up to the constant $2C_F$,
so $\eta^{(2)}_q=\eta^{(1)}_q$ at this order, and the gluon column
gives
\begin{equation}
    \eta^{(2)}_g(\mu) = 1 + a_s(\mu)\Big(\frac{2N_fT_F}{3} - \frac{\pi^2}{6}\,C_A\Big)
    + O(a_s^2),
    \label{eq:tmd_eta_g}
\end{equation}
about $0.90$ at $\alpha_s=0.35$ for $N_f=4$. With
$\eta^{(1)}_q=\eta^{(2)}_q\equiv\eta_q$ the first two lines of
Eq.~\eqref{eq:tmd_admissible_all} collapse, and the normalized scheme
becomes a rescaling of each species,
\begin{equation}
    f_i(x;\mu) = \frac{1}{\eta_i(\mu)}
    \int_{|\boldsymbol k_T|<\mu}\dd^2\boldsymbol k_T\,
    f_i(x,\boldsymbol k_T;\mu,\mu^2),
    \label{eq:tmd_admissible}
\end{equation}
with $\eta_q$ from Eq.~\eqref{eq:tmd_eta_q1} and $\eta_g=\eta^{(2)}_g$
from Eq.~\eqref{eq:tmd_eta_g}. We have checked numerically that
Eq.~\eqref{eq:tmd_admissible_all} with the one-loop kernels satisfies
both covector conditions exactly, and that
Eq.~\eqref{eq:tmd_admissible} does so up to terms of order $a_s^2$.

On the TMD side the normalization is nothing exotic. A factorization
theorem only fixes the product of the hard factor and the TMDs, so a
constant factor common to all quark TMDs, or attached to the gluon TMD,
is a convention, and one is free to move it into the hard factor. The
sum rules pick the convention. With the quark TMD normalized by
$1/\eta_q$, its integral over transverse momentum up to the scale is
the collinear density with the right valence numbers, and the integral
relation holds with no correction at all beyond the normalization.
This is the prescription that keeps both readings at once, the
parton-model reading of the collinear densities through the sum rules
and the probabilistic reading of the TMD as a density in transverse
momentum whose integral is the collinear density. To be clear, there
is nothing wrong with the unnormalized cutoff scheme. It is a
consistent scheme, as every DGLAP-covariant transformation is, and
Sec.~\ref{sec:covectors} already made the point that giving up a sum
rule is not an inconsistency. What this section adds is the
identification of the scheme that relates TMDs and collinear densities
while keeping the whole partonic interpretation, and the fact that it
is the unnormalized scheme times a constant matrix fixed by three
numbers.

Furthermore, one could instead move the cutoff to
$\lambda\mu$, which adds logarithms of $\lambda$ to the moments and can
be tuned to make $c_1$ vanish, but a zero obtained this way does not
survive the next order, while the normalization works at every order. And
if the scales of the TMD are left independent of the cutoff, the
integrated density does not even obey an equation of the DGLAP form,
and no finite transformation can restore it \cite{delRio:2024}, so it
falls outside Definition~\ref{def:admissible} from the start.

\part{Solution of DGLAP by scheme transformation}
\label{part:two}

In this second part we use the scheme transformations to solve the DGLAP equations and
to quantify missing higher-order uncertainties. It can be read
independently of Part~\ref{part:one}. The main inputs we borrow from
there are the transformation law \eqref{eq:gauge_transformation} of the
kernels, the boundary condition \eqref{eq:asymptotic_identity}, and the
block decomposition \eqref{eq:irrep_decomposition} of flavor space. A
few remarks along the way compare with the classification, and
Sec.~\ref{sec:scheme_variation} returns to it deliberately.

\section{Gauging the evolution away}
\label{sec:solution}

In Part~\ref{part:one} we classified the transformations that preserve
the symmetries and the sum rules. We now
show that the same gauge structure also solves the evolution. The logic
was already visible at the end of Sec.~\ref{sec:other_moments}.
Pointwise in $z$, DGLAP is a connection in the single variable $\mu$,
and a one-dimensional connection is pure gauge, so finding the
transformation that gauges it away is the same as solving DGLAP. One
caveat separates this part from the classification: a transformation
that absorbs the evolution cannot approach the identity at large $\mu$,
so the schemes of this section violate the boundary condition
\eqref{eq:asymptotic_identity} deliberately. They are computational
devices, not admissible schemes, and, consistently, the injectivity
argument of Sec.~\ref{sec:injectivity} does not apply to them. The ``same
evolution'' freedom that was eliminated there by the boundary condition
will reappear here as the choice of fundamental matrix.

The representation we arrive at is equivalent term by term to the
standard Mellin-space solution, as Appendix~\ref{sec:dressed} shows.
What it adds is structural. It makes the radius of convergence and the
window in $z$ explicit, it explains the origin of the resonant
moments, it separates at the level of single coefficients what a given
order knows from what it does not, and it links the missing higher
orders back to Part~\ref{part:one} through scheme variation restricted
to the admissible transformations.

\subsection{The no-evolution and diagonal schemes}
\label{sec:no_evolution}

Let $\Kc_0$ gauge the connection away entirely, $\gamma^{(0)}=0$, where
the superscript is a scheme label as in
Eq.~\eqref{eq:gauge_transformation} and not a perturbative order. By
Eq.~\eqref{eq:gauge_transformation} this means
\begin{equation}
    \frac{\dd\Kc_0}{\dd\ln\mu^2} = -\Kc_0\,\gamma ,
    \label{eq:no_evolution_condition}
\end{equation}
and in the new scheme the densities do not run at all. Undoing the
transformation at both ends turns the evolution factor into a plain
matrix product,
\begin{equation}
\begin{gathered}
    \overline f(z;\mu) = E_\gamma(z;\mu,\mu_0)\,\overline f(z;\mu_0),\\
    E_\gamma(z;\mu,\mu_0) = \Kc_0^{-1}(z;\mu)\,\Kc_0(z;\mu_0) ,
\end{gathered}
    \label{eq:E_from_K0}
\end{equation}
and no path ordering survives. Note that this is the same evolution
operator $E_\gamma$ as in Eq.~\eqref{eq:adjoint_transport}, now built
without path-ordered exponentials. Of course,
Eq.~\eqref{eq:no_evolution_condition} is exactly as hard as DGLAP itself.

The gain comes from gauging in two steps. First transform to a scheme
where the connection is merely diagonal,
\begin{equation}
    \Kc_D\,\gamma\,\Kc_D^{-1}
    + \frac{\dd\Kc_D}{\dd\ln\mu^2}\Kc_D^{-1} = \overline D
    \quad \text{diagonal},
    \label{eq:diagonal_scheme}
\end{equation}
because an abelian connection needs no path ordering either,
\begin{equation}
    \overline f(\mu) = \Kc_D^{-1}(\mu)\,
    \exp\!\Big(\int_{\mu_0^2}^{\mu^2}\dd\ln\mu'^2\,\overline D\Big)\,
    \Kc_D(\mu_0)\,\overline f(\mu_0).
\end{equation}
The condition
\eqref{eq:diagonal_scheme}, with the derivative term included, also
appears in small-$x$ resummation, Eq.~(2.8) of Ref.~\cite{Bonvini:2016},
where it is observed that the derivative term is suppressed by
$a_s\beta_0$ relative to the similarity term and can be treated
perturbatively. Here we keep it exactly. Second, the block structure of
Sec.~\ref{sec:symmetries} is also the natural basis for solving the
evolution. In the decomposition \eqref{eq:irrep_decomposition} every
sector carries a single scalar, and is therefore abelian, except the
$2\times2$ singlet-gluon block. The whole problem reduces to diagonalizing one
$2\times2$ connection
$\gamma = \big(\begin{smallmatrix}\gamma_{qq} & \gamma_{qg}\\
\gamma_{gq} & \gamma_{gg}\end{smallmatrix}\big)$
acting on $(\overline\Sigma_+, \overline g)$. A scheme without
quark-gluon mixing has also been advocated on physical grounds, for
observables that are not fully inclusive, in Ref.~\cite{Oliveira:2013}.
Here we use it purely as a computational device.

With the normalization $\Kc_0(\mu_0)=\Id$ the no-evolution density is
the input-scale density,
$\overline f^{(0)}(z) = \Kc_0(\mu)\overline f(\mu) = \overline f(\mu_0)$.
The transformation lies in the commutant and fixes every sum-rule
covector, so it satisfies every constraint of Part~\ref{part:one}
except the boundary condition \eqref{eq:asymptotic_identity}, the
independent physical input that Sec.~\ref{sec:injectivity} showed
removes the ``same evolution'' orbits.

\subsection{Reduction to a Riccati equation}
\label{sec:riccati}

Write $\Kc_D$ with entries $\Kc_{qq},\Kc_{qg},\Kc_{gq},\Kc_{gg}$ and
demand that the off-diagonal parts of Eq.~\eqref{eq:diagonal_scheme}
vanish. The condition from the first row is (prime
$=\dd/\dd\ln\mu^2$)
\begin{equation}
    \Kc_{qq}' = \frac{\gamma_{qg}}{\Kc_{qg}}\Kc_{qq}^2
    + \left(\frac{\Kc_{qg}'}{\Kc_{qg}} - \gamma_{qq}
    + \gamma_{gg}\right)\Kc_{qq} - \Kc_{qg}\,\gamma_{gq} ,
    \label{eq:Riccati}
\end{equation}
which is a Riccati equation, with its mirror image
($q\leftrightarrow g$) on the second row. The standard substitution
$\Kc_{qq} = -(\Kc_{qg}/\gamma_{qg})\,u_{qq}'/u_{qq}$ linearizes it,
\begin{equation}
\begin{gathered}
    u_{qq}'' - \Lambda_{qq}\, u_{qq}' + \Phi\, u_{qq} = 0,\\
    \Lambda_{qq} = \gamma_{gg} - \gamma_{qq}
    + \frac{\gamma_{qg}'}{\gamma_{qg}},
    \qquad
    \Phi = -\gamma_{qg}\gamma_{gq} ,
\end{gathered}
    \label{eq:u_equation}
\end{equation}
with $\Lambda_{gg}=\gamma_{qq}-\gamma_{gg}+\gamma_{gq}'/\gamma_{gq}$ for
the mirror equation. Notice that the free function $\Kc_{qg}$ has dropped
out. The coefficients are built from the anomalous dimensions alone, so
Eq.~\eqref{eq:u_equation} is the gauge-invariant content of the problem.
The residual freedom only moves the eigenvalues
$\overline D_{qq} = \Kc_{qg}'/\Kc_{qg} + \gamma_{gg} - u_{qq}'/u_{qq}$
around, which is the abelian gauge freedom left inside the diagonal
scheme. Choosing $\overline D = 0$ lands on the no-evolution scheme, with
(normalizations set to one)
\begin{equation}
\begin{gathered}
    \Kc_0(\mu) =
    \begin{pmatrix}
        -\dfrac{u_{qq}'}{\gamma_{qg}}\,\omega_g
        & u_{qq}\,\omega_g\\[8pt]
        u_{gg}\,\omega_q
        & -\dfrac{u_{gg}'}{\gamma_{gq}}\,\omega_q
    \end{pmatrix},\\[4pt]
    \omega_{q,g}(\mu) \equiv
    e^{-\int_{\mu_0^2}^{\mu^2}\dd\ln\mu'^2\,\gamma_{qq,gg}(\mu')},
\end{gathered}
    \label{eq:K0_explicit}
\end{equation}
and the exact all-orders evolution factor is the $2\times2$ matrix
product \eqref{eq:E_from_K0}.

Two facts turn the construction from a formal statement into a
practical one.
First, only one second-order equation actually needs solving. The generalized Wronskian
$W \equiv u_{qq}'u_{gg}' - \gamma_{qg}\gamma_{gq}\,u_{qq}u_{gg}$ obeys
$\left(W/\gamma_{qg}\gamma_{gq}\right)'=0$, which is Abel's theorem for
the pair of adjoint equations, so $u_{gg}$ follows from $u_{qq}$ by a
single quadrature.

Second, the integration constants this leaves open,
together with every normalization choice made above, amount to a constant
left factor $\Kc_0 \to M\Kc_0$. These are precisely the ``same
evolution'' equivalence classes of Sec.~\ref{sec:injectivity}, which are
harmless here since $M$ cancels identically in
$E_\gamma = \Kc_0^{-1}(\mu)\Kc_0(\mu_0)$. The same Abel relation gives
$\det\Kc_0 \propto e^{-\int\dd\ln\mu'^2\,\mathrm{Tr}\,\gamma}$, which
never vanishes if it is nonzero at the input scale, so the construction
cannot become singular along the evolution.

The determinant can still be
zero at the input scale itself, because each row of $\Kc_0$ behaves at
small coupling as a power of $a_s$ governed by one of the two
eigenvalues of $\gamma^{[1]}$, with the choice of solutions $u_{qq}$ and
$u_{gg}$ deciding which one for each row. If the two rows carry the same
eigenvalue they are proportional to each other and $W$ is zero at every
scale. Choosing opposite eigenvalues for the two rows avoids this, and
it is also the physical choice, since the exact evolution contains both
modes and $\Kc_0$ must carry both to reproduce it.

\subsection{Series solution at any order}
\label{sec:frobenius}

Trading $\ln\mu^2$ for the coupling
($\dd/\dd\ln\mu^2 = \beta\,\dd/\dd a_s$) puts
Eq.~\eqref{eq:u_equation} in the form
\begin{equation}
\begin{gathered}
    \ddot u + \frac{\lambda(a_s)}{a_s}\,\dot u
    + \frac{\phi(a_s)}{a_s^2}\, u = 0 ,\\
    \lambda \equiv a_s\,\frac{\dot\beta-\Lambda}{\beta},
    \qquad
    \phi \equiv -a_s^2\,\frac{\gamma_{qg}\gamma_{gq}}{\beta^2},
\end{gathered}
    \label{eq:frobenius_form}
\end{equation}
with dots denoting $\dd/\dd a_s$, $\Lambda$ the coefficient of
Eq.~\eqref{eq:u_equation} for the row considered, and $\lambda,\phi$
rational functions of $a_s$. Their polynomial degree grows with the
perturbative order but the structure of the equation does not. Going to
N$^k$LO only lengthens the polynomials. The point $a_s=0$ is a regular
singular point, so a Frobenius solution $u = a_s^{R}\sum_i A_i\, a_s^i$
always exists, with indices fixed once and for all by the leading-order
anomalous dimensions,
\begin{equation}
    R_{qq} = \frac{\gamma^{[1]}_{qq}-\gamma^{[1]}_{gg}
    \pm\sqrt{\big(\gamma^{[1]}_{qq}-\gamma^{[1]}_{gg}\big)^2
    + 4\,\gamma^{[1]}_{qg}\gamma^{[1]}_{gq}}}{2\beta_0} ,
    \label{eq:frobenius_indices}
\end{equation}
and similarly for $R_{gg}$ with $q\leftrightarrow g$.

It is worth checking the power counting behind these statements. In QCD
$\beta$ starts at $O(a_s^2)$, while $\gamma$, and with it $\Lambda$,
starts at $O(a_s)$. This includes the logarithmic-derivative term of
$\Lambda$, since
$\gamma_{qg}'/\gamma_{qg} = \beta\,\dd\ln\gamma_{qg}/\dd a_s
= -\beta_0\,a_s + O(a_s^2)$, where the $1/a_s$ coming from the logarithm
is compensated by the $a_s^2$ of $\beta$. The numerator and the
denominator of $\lambda$ therefore both start at $O(a_s^2)$, and
$\lambda$ and $\phi$ are finite and nonzero at $a_s=0$,
\begin{equation}
\begin{gathered}
    \lambda(a_s=0)\equiv l_0 = 1 - \frac{\gamma^{[1]}_{qq}-\gamma^{[1]}_{gg}}{\beta_0},\\
    \phi(a_s=0)\equiv p_0 = -\frac{\gamma^{[1]}_{qg}\gamma^{[1]}_{gq}}{\beta_0^{2}}
\end{gathered}
    \label{eq:l0_p0}
\end{equation}
(for the $qq$ row, with $q\leftrightarrow g$ for the other). This is
precisely what makes $a_s=0$ a regular singular point, since the
coefficients $\lambda/a_s$ and $\phi/a_s^2$ carry poles of order exactly
one and two, and the indicial polynomial $F(x)=x(x-1)+l_0x+p_0$ built
from Eq.~\eqref{eq:l0_p0} reproduces Eq.~\eqref{eq:frobenius_indices},
as it must. At leading order $\lambda$ and $\phi$ are constants, the
equation is of Cauchy-Euler type, $u = a_s^{R_\pm}$, and the standard
closed-form LO evolution factor is recovered.

The series is fully explicit at any order. Expanding the coefficient
functions as $\lambda(a_s)=\sum_{k\geqslant0}l_k\,a_s^k$ and
$\phi(a_s)=\sum_{k\geqslant0}p_k\,a_s^k$ (the $l_k,p_k$ are fixed
algebraic combinations of the $\gamma^{[j]}$ and $\beta_j$ of the chosen
truncation) and inserting the ansatz
$u = a_s^{R}\sum_{i\geqslant0}A_i\,a_s^i$, the coefficients obey the
recursion
\begin{equation}
\begin{gathered}
    A_0 = 1,\\
    A_m = -\frac{1}{F(R+m)}\sum_{k=1}^{m}
    \big[\,l_k\,(R+m-k) + p_k\,\big]\,A_{m-k},
\end{gathered}
    \label{eq:frobenius_recursion}
\end{equation}
where $F(x)\equiv x(x-1)+l_0\,x+p_0=(x-R_+)(x-R_-)$ is the indicial
polynomial whose roots are the indices \eqref{eq:frobenius_indices}. The
general solution is $u=c_+u_++c_-u_-$ built on the two indices, and the
constants $c_\pm$ join the fundamental-matrix freedom that cancels in
the evolution factor. Since $F(R_\pm+m)=m\,\big(m\pm(R_+-R_-)\big)$, the
recursion for the smaller index fails when the two indices differ by a
positive integer. At those special Mellin moments the second solution
develops a logarithm and the coefficients of the smaller index are
singular, while the evolution operator, in which the two solutions
combine, stays regular. These are the moments where the map of
Eq.~\eqref{eq:injectivity_operator} degenerates, and the singular
coefficients are a feature of the representation and not of the
evolution.

On the physical domain of small positive $a_s$
the coefficients $\lambda$ and $\phi$ are perfectly regular. Their poles
lie elsewhere in the complex $a_s$ plane, at the zeros of the truncated
$\beta$ polynomial (at NLO, $a_s=-\beta_0/\beta_1$, on the negative real
axis) and of the truncated off-diagonal anomalous dimensions (at NLO,
$a_s=-\gamma^{[1]}_{qg}/\gamma^{[2]}_{qg}$). All of them are spurious.
The QCD $\beta$ function has no zero in the perturbative regime and the
evolution factor is perfectly regular at these points, which are
artifacts of truncating the rational coefficients and not of the
physics.

Nevertheless they control the series, because the radius of
convergence of a power series about $a_s=0$ is the distance to the
nearest singularity anywhere in the complex plane, regardless of whether
physics ever visits it. The Taylor series of $1/(1+x^2)$, for instance,
has radius one on the perfectly real-analytic real line because of the
poles at $\pm i$. The series therefore converges for
$|a_s| < \rho(z) = \min\big(|\beta_0/\beta_1|,\,
|\gamma^{[1]}_{qg}/\gamma^{[2]}_{qg}|,\,
|\gamma^{[1]}_{gq}/\gamma^{[2]}_{gq}|\big)$ at NLO, with the analogous
higher zeros beyond. Numerically the $\beta$ zero sits at
$|a_s|\approx0.16$, that is $\alpha_s\approx2$, well beyond the deepest
infrared coupling one ever evolves from, so it never limits anything in
practice.

Only the $z$-dependent family matters, since it can approach
the origin at particular points of the Mellin inversion contour, where
convergence must be monitored moment by moment. Even then, the radius
limits the series representation and not the solution. The function
$u(a_s)$ continues analytically past $\rho(z)$, and one may integrate
the ODE directly or re-expand about a different point.

It is instructive to compare with the standard Mellin-space solution,
implemented in \textsc{Pegasus} \cite{Vogt:2005} and available among
the evolution modes of Ref.~\cite{Candido:2022}. That method stays
with the first-order matrix equation and writes the evolution operator
as the leading-order factor dressed on both sides by a power series of
matrices, truncated at the order of the kernels, with prescriptions
that differ beyond it. The dressing matrices obey commutator
recursions whose denominators vanish at the moments where the indices
of Eq.~\eqref{eq:frobenius_indices} differ by an integer, so the
resonances above appear there as singular coefficients, which the
indicial structure above accounts for. The two expansions agree term by term since the dressing matrices applied to the leading-order
eigenvectors reproduce the Frobenius coefficients. The difference between the two only
lies in what is kept and not in what is computed.
In Appendix~\ref{sec:dressed} we derive the formal identification. Iterating the
dressing to all orders converges, inside the radius $\rho(z)$, to the same
resummed solution that the Frobenius series delivers directly, and the
series keeps explicit what a fixed-order truncation does not display,
the radius itself, the window in $z$, and the account of which
coefficients are exact at a given order that drives
Sec.~\ref{sec:mhou}. None of this is a claim
of practical advantage. The comparison explains the structure of the
dressed expansion, its singular denominators included, and validates
it as the fixed-order truncation of a convergent series.

A further analytic route that approximates the evolution operator by a closed
product of matrix exponentials obtained from the Magnus expansion
\cite{Simonelli:2024} has been studied. The main feature of the present construction is
that, just as the admissible
schemes were classified once and for all, the evolution is solved by a
single equation whose form is order independent. Perturbative accuracy enters only through the length of the
two polynomial coefficient lists.

We have verified the whole construction numerically at NLO, using the
$\overline{\rm MS}$ anomalous dimensions as implemented in
Ref.~\cite{Candido:2022}, with
$N_f=4$ held fixed, no heavy-quark thresholds, and the coupling running
from $\alpha_s(\mu_0)=0.35$ down to $\alpha_s(\mu)=0.118$, which
corresponds to $\mu_0\approx1.9$~GeV and $\mu=M_Z$ when the coupling is
anchored at the $Z$-boson mass. This is a mathematical check of the
construction. We do not claim any phenomenological result here.

\begin{figure}[!t]
    \centering
    \includegraphics[width=\columnwidth]{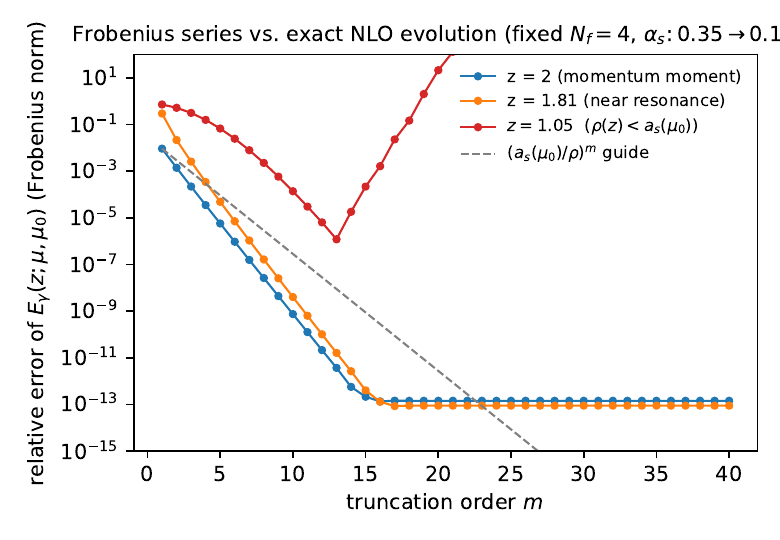}
    \caption{Relative error, in the Frobenius norm, of the singlet-gluon
    block of the evolution factor, the $2\times2$ matrix acting on
    $(\overline\Sigma_+,\overline g)$, built from the
    Frobenius series, as a function of the number $m$ of terms kept in
    the recursion \eqref{eq:frobenius_recursion}, against a direct
    high-accuracy integration of the NLO evolution with $N_f=4$ held
    fixed and $\alpha_s$ running from $0.35$ to $0.118$. At $z=2$, and
    near the degenerate moment $z\simeq1.80$, the error falls at the
    geometric rate $\big(a_s(\mu_0)/\rho(z)\big)^m$ set by the
    convergence radius until it reaches the accuracy of the integrator.
    At $z=1.05$ the radius lies below $a_s(\mu_0)$ and the truncated
    series is asymptotic, with the error reaching a minimum and then
    growing.}
    \label{fig:convergence}
\end{figure}

In Figure~\ref{fig:convergence} we compare the singlet-gluon block of the
evolution factor, the $2\times2$ matrix acting on
$(\overline\Sigma_+,\overline g)$ and the only sector where the series
construction is nontrivial, built from the truncated series against a
direct high-accuracy integration of the evolution equation. Here $m$
denotes the number of terms kept in the recursion
\eqref{eq:frobenius_recursion} and $\Kc_0$ is assembled as in
Eq.~\eqref{eq:K0_explicit}, with the two rows built on opposite
eigenvalues as discussed in Sec.~\ref{sec:riccati}, and the error is
the Frobenius norm of the difference divided by the Frobenius norm of
the exact block, a measure that captures all four entries at once. At $z=2$, and also
close to the degenerate moment $z\simeq1.80$, the relative error falls
geometrically at the predicted rate
$\big(a_s(\mu_0)/\rho(z)\big)^m$ until it reaches the accuracy of the
integrator, and the Wronskian combination $W/(\gamma_{qg}\gamma_{gq})$
of Sec.~\ref{sec:riccati} stays constant to the same accuracy. At
$z=1.05$, where $\rho(z)<a_s(\mu_0)$, the truncated series behaves as an
asymptotic one, with the error decreasing until $m\approx13$ and growing
afterwards, which confirms the radius formula from the divergent side as
well.

The radius shrinks to zero as $z\to1^{+}$ because the NLO
off-diagonal anomalous dimensions have a pole at $z=1$. It also falls
slowly at large $|z|$, since the NLO off-diagonal anomalous dimensions
grow by a logarithm of $z$ relative to LO, so for
$\alpha_s(\mu_0)=0.35$ the series converges in the window
$1.4\lesssim|z|\lesssim21$, closed at both ends, although the closure
is soft. Outside the window the truncated series
still behaves as an asymptotic expansion, as the $z=1.05$ curve of
Fig.~\ref{fig:convergence} shows, and truncating at the order of
minimum error remains accurate there, while far outside one integrates
Eq.~\eqref{eq:u_equation} directly at the affected contour points. On a
Mellin inversion contour the two closures therefore matter only for the
reconstruction at very small and very large $x$.

\begin{figure}[!t]
    \centering
    \includegraphics[width=\columnwidth]{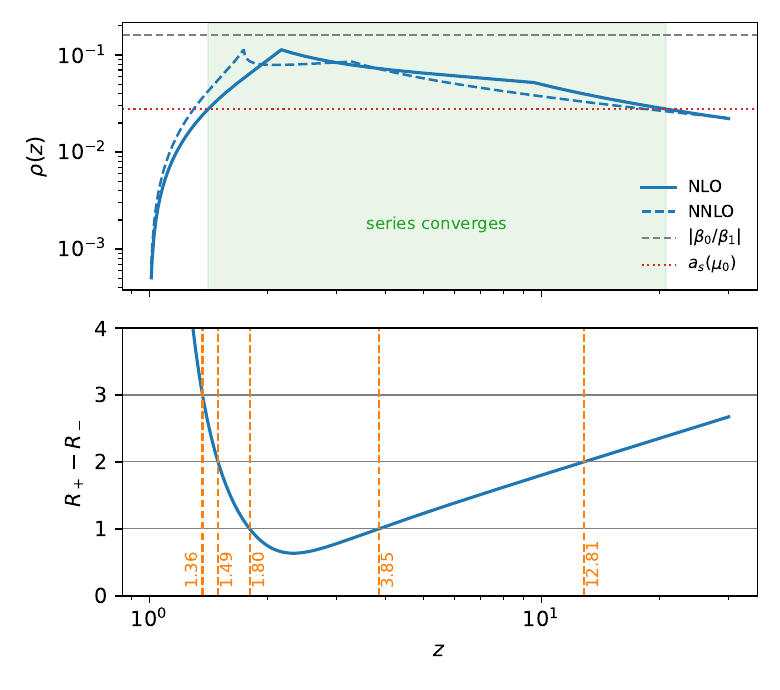}
    \caption{The NLO convergence radius $\rho(z)$ of the Frobenius
    series (upper panel) and the index difference $R_+-R_-$ of
    Eq.~\eqref{eq:frobenius_indices} (lower panel) on the real axis for
    $N_f=4$. The shaded region marks the window where the series
    converges for $\alpha_s(\mu_0)=0.35$, and the dashed curve is the
    NNLO radius, which moves the window only mildly. The radius drops to zero as
    $z\to1^{+}$, because the NLO off-diagonal anomalous dimensions have
    a pole there, and falls logarithmically at large $z$, so the window
    is closed at both ends. The index difference equals a positive
    integer at the degenerate moments, marked by the vertical lines,
    and grows without bound toward $z=1$.}
    \label{fig:radius}
\end{figure}

In Figure~\ref{fig:radius} we collect $\rho(z)$ and the index difference
$R_+-R_-$ on the real axis. For $N_f=4$ the difference has a minimum
near $z\simeq2.33$, equals one at $z\simeq1.80$ and $z\simeq3.85$, and
equals two at $z\simeq1.49$ and again at $z\simeq12.8$. It grows
without bound toward $z=1$, so the degenerate moments accumulate there,
and it climbs logarithmically at large $z$, where it meets the higher
integers one after the other.

The same code confirms the
transport of the sum rules along the evolution, since the momentum
covector at $z=2$ and the valence numbers at $z=1$ are preserved by the
computed evolution factors to the accuracy of the input anomalous
dimensions.

We have repeated all of these checks at NNLO, with the three-loop
anomalous dimensions of Refs.~\cite{Moch:2004,Vogt:2004} as implemented
in Ref.~\cite{Candido:2022}, and the structure is unchanged. Only the
coefficient lists $l_k$ and $p_k$ grow longer, the recursion and the
assembly of $\Kc_0$ stay the same, and the truncated series again
reaches the accuracy of the integrator at the geometric rate set by the
NNLO radius. The indices \eqref{eq:frobenius_indices} depend only on
$\gamma^{[1]}$ and $\beta_0$, so the degenerate moments are the same at
every order. The radius does change, since its singularities move with
the truncation, but only mildly, and the convergence window shifts from
$1.4\lesssim|z|\lesssim21$ at NLO to $1.3\lesssim|z|\lesssim18$ at
NNLO, as the dashed curve of Fig.~\ref{fig:radius} shows.

\section{Missing higher orders in the evolution}
\label{sec:mhou}

\subsection{What is and what is not known at a given order}
\label{sec:mhou_known}

The Frobenius representation makes it possible to say precisely which
part of the evolution is unknown once the perturbative series is
truncated. At N$^{\,n-1}$LO one possesses
$\gamma^{[1]},\dots,\gamma^{[n]}$ and $\beta_0,\dots,\beta_{n-1}$. The
subtlety is that promoting the accuracy by one order does not merely
append new terms to the coefficient functions of
Eq.~\eqref{eq:frobenius_form}. A direct computation gives
\begin{equation}
\begin{gathered}
    \lambda_{(n+1)}(a_s)-\lambda_{(n)}(a_s) = O(a_s^{\,n}),\\
    \phi_{(n+1)}(a_s)-\phi_{(n)}(a_s) = O(a_s^{\,n}),
\end{gathered}
    \label{eq:order_shift}
\end{equation}
where the subscript counts the truncation order. In other words, the new
order corrects coefficients that are already present. Consequently the
expansion coefficients $l_k,p_k$ of Eq.~\eqref{eq:frobenius_recursion}
with $k<n$ are exact, while those with $k\geqslant n$ are only partially
known, in the sense that they contain a definite contribution from the
known orders plus an unknown shift from the orders beyond. Feeding this
into the recursion we obtain
\begin{equation}
\begin{gathered}
    A_i^{\,\text{N}^{n-1}\text{LO}} = A_i^{\,\text{exact}}
    \quad (0\leqslant i\leqslant n-1),\\
    A_n^{\,\text{N}^{n-1}\text{LO}} \neq A_n^{\,\text{exact}}.
\end{gathered}
    \label{eq:which_A_exact}
\end{equation}
That is, $A_0$ is fully determined at LO, $A_0$ and $A_1$ at NLO, and
each new order determines exactly one more coefficient of the series.

One immediate consequence is a genuine protection. The indices $R_\pm$ are
built from $l_0$ and $p_0$ alone, which by Eq.~\eqref{eq:order_shift}
never receive corrections, so $R$ is fixed at leading order, exactly, to
all orders. Since $u\to A_0\,a_s^{R}$ as $a_s\to0$, the asymptotic
($\mu\to\infty$) behavior of the evolution carries no missing-order
uncertainty whatsoever. A prescription that assigns one to it, as a
uniform scale variation does, is conservative there. The opposite
consequence is a caveat. The relative truncation error is
\begin{equation}
    \frac{u^{\,\text{N}^{n}\text{LO}}-u^{\,\text{N}^{n-1}\text{LO}}}
    {u^{\,\text{N}^{n-1}\text{LO}}}
    = \frac{\delta A_n}{A_0}\,a_s^{\,n} + O(a_s^{\,n+1}),
    \label{eq:truncation_scaling}
\end{equation}
suppressed by $a_s^{n}$. At NNLO the first
contaminated coefficient is $A_3$, precisely because
raising the order corrects existing coefficients instead of only
appending new ones.

\subsection{The leading missing-order shift}
\label{sec:mhou_scale}

To obtain the leading shift in closed form, denote by
$\delta l_n,\delta p_n$ the changes of the first contaminated
coefficients when the next order becomes available. Since the exact
coefficients $A_0,\dots,A_{n-1}$ and the exact $l_k,p_k$ with $k<n$ do
not move, the recursion \eqref{eq:frobenius_recursion} gives
\begin{equation}
    \frac{\delta A_n}{A_0}
    = -\,\frac{R\,\delta l_n + \delta p_n}{F(R+n)},
    \quad
    F(R_\pm+n) = n\big(n\pm(R_+-R_-)\big),
    \label{eq:master_shift}
\end{equation}
with $\delta l_n,\delta p_n$ linear in the unknowns $\gamma^{[n+1]}$ and
$\beta_n$.

This is a compact parametrization of the theory
uncertainty on the evolution. At each Mellin moment the entire
sensitivity to the missing order is carried by two numbers per row, and
any future determination of $\gamma^{[n+1]}$, even a partial one such as
a few moments or a large or small $z$ limit, propagates through
Eq.~\eqref{eq:master_shift} without further work.

The shift of the
observable evolution follows from the logarithmic derivative,
\begin{equation}
    \delta\!\left(\frac{\dd\ln u}{\dd\ln\mu^2}\right)
    = -\left(n\,\beta_0\,\frac{\delta A_n}{A_0}
    + R\,\beta_n\right) a_s^{\,n+1}
    + O(a_s^{\,n+2}),
    \label{eq:delta_logderiv}
\end{equation}
where the first term carries the shifted coefficient $A_n$ and the
second comes from the new coefficient of the $\beta$ function acting on
the leading power $a_s^{R}$. Both here and in
Eq.~\eqref{eq:master_shift} the statement holds for each of the two
solutions separately, with $R$ the corresponding index $R_\pm$ of
Eq.~\eqref{eq:frobenius_indices}. The shift is one power of $a_s$ softer
than $\delta u/u$ and is the quantitative version of the protection
discussed above. The size of the
correction is therefore parametric and computable. The missing-order
band on the evolution factor has relative size $a_s^{\,n}(\mu_0)$ times
a $z$-dependent coefficient that is fixed once the unknowns are
assigned their natural size. The constant in
$|\gamma^{[n+1]}|\lesssim O(1)\times|\gamma^{[n]}|$ should be
calibrated on the growth of the known orders, which in this
normalization is a factor of eight to ten for the non-singlet but only
about two for the singlet block at $z=2$.

One region of the Mellin plane requires care, the region where the
denominator of Eq.~\eqref{eq:master_shift} vanishes. On the branch $R_-$
this happens when the two indices differ by a positive integer, which is
the degenerate Frobenius case already met below
Eq.~\eqref{eq:frobenius_recursion}. The condition is satisfied at
real, physically sampled moments. For $N_f=4$ and $n=1$ it happens at
$z\simeq1.80$ and $z\simeq3.85$, the latter in the middle of a typical
inversion contour, and for $n=2$ at $z\simeq1.49$ and $z\simeq12.8$,
the same moments at which Eq.~\eqref{eq:injectivity_operator}
degenerates. There the shift of one branch diverges while the shift of
the evolution operator does not, since the divergence cancels between
the two branches. We have checked that the exact change of the
evolution operator from LO to NLO passes smoothly through the first
pair and from NLO to NNLO through the second. A band built from the
full operator therefore shows no structure at these moments, while one
built from Eq.~\eqref{eq:master_shift} on a single branch would show a
spurious peak, so in their neighborhood the logarithmic branch, or the
operator itself, must be used.

\begin{figure}[!t]
    \centering
    \includegraphics[width=\columnwidth]{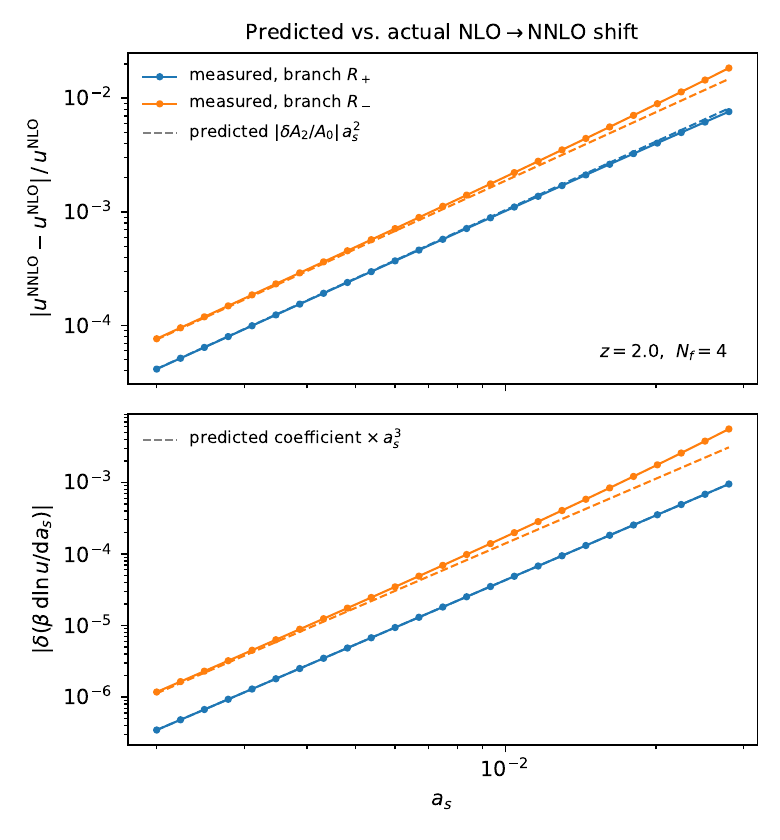}
    \caption{The change of the Frobenius solution (upper panel) and of
    its logarithmic derivative (lower panel) between NLO and NNLO at
    $z=2$ with $N_f=4$. The dashed lines are the predictions of
    Eqs.~\eqref{eq:master_shift} and \eqref{eq:delta_logderiv}
    evaluated with the true $\gamma^{[3]}$ and $\beta_2$, so the powers
    $a_s^{2}$ and $a_s^{3}$ fix the slopes and the predicted
    coefficients fix the offsets. The measured points follow the
    predictions and peel away at larger coupling by the expected one
    extra power of $a_s$, visibly on the $R_-$ branch and mildly on
    $R_+$.}
    \label{fig:mhou}
\end{figure}

Because the NNLO anomalous dimensions exist, the case $n=2$ can be
closed completely. We treat the NNLO truncation as the truth and the
NLO one as the working order, so that $\gamma^{[3]}$ and $\beta_2$ play
the role of the unknowns, and we compare every formula of this section
with the actual change from NLO to NNLO. In this setting
$\lambda_{(3)}-\lambda_{(2)}$ and $\phi_{(3)}-\phi_{(2)}$ start exactly
at order $a_s^{2}$, the coefficients $A_0$ and $A_1$ do not move while
$A_2$ does, and Eq.~\eqref{eq:master_shift} reproduces the shift of
$A_2$ exactly on both branches, since for the first
contaminated coefficient it is an exact identity and not only a leading
approximation. The measured changes of $u$ and of its logarithmic
derivative then approach Eqs.~\eqref{eq:truncation_scaling} and
\eqref{eq:delta_logderiv} as $a_s\to0$ at the expected rate, as
Fig.~\ref{fig:mhou} shows. The
missing-order formulas therefore predict the actual next order, and not
only its parametric size. At the sum-rule moments the true shift is
also much smaller than an uncorrelated estimate suggests, since the
covector conditions of Part~\ref{part:one} constrain $\gamma^{[n+1]}$
there as well, so the protected moments are protected from missing
orders too. At the physical input coupling the leading
term is, however, not numerically dominant on both branches. At
$a_s(\mu_0)\approx0.028$ the next correction changes the shift of the
logarithmic derivative by a few percent on the $R_+$ branch but by a
sizable fraction of the leading term on $R_-$, so at low scales the
leading formula fixes the structure and the order of magnitude of a
missing-order band, while a precise band on the subleading branch
requires the $O(a_s^{\,n+2})$ term as well.

\subsection{Scheme variation as a sum-rule-preserving uncertainty estimate}
\label{sec:scheme_variation}

The series analysis above quantifies the truncation uncertainty of the
evolution at fixed scheme. Scheme variation probes the same ignorance
from a different direction, and it does so cleanly only if one varies
within admissible schemes. Otherwise the variation mixes a violation of
the sum rules into the perturbative uncertainty.

Part~\ref{part:one} tells us what an admissible variation looks like at
every moment, not only at $z=1,2$ where Eq.~\eqref{eq:final_group}
applies. At working order N$^{\,n-1}$LO the variation must leave the
known kernels alone and move the first missing one, so the general
family is
\begin{equation}
    \Kc(z;\mu) = \Id + \kappa(z)\,a_s^{\,n}(\mu) + O(a_s^{\,n+1}),
    \label{eq:mhou_family_general}
\end{equation}
with $\kappa(z)$ an analytic matrix subject to three requirements. It
must lie in the commutant \eqref{eq:commutant_arena} at every $z$, so it
consists of a $2\times2$ block on $\{\Sigma_+,g\}$ and three scalars. It
must be real at real $z$ by the Schwarz reflection of
Sec.~\ref{sec:mellin}(iii). And it must respect the sum rules at the two
special moments, so at $z=2$ the columns of its singlet-gluon block sum
to zero and at $z=1$ its $C$-odd scalars vanish. In $x$ space $\kappa$
is a convolution kernel, and genuine scheme differences, the DIS, Monte
Carlo and positivity schemes for instance, are of this $z$-dependent
kind \cite{Delorme:2025}. Inserting the family into the transformation
law \eqref{eq:gauge_transformation} gives the shift it induces on the
first missing kernel,
\begin{equation}
    \delta\gamma^{[n+1]} = \big[\kappa,\gamma^{[1]}\big] - n\,\beta_0\,\kappa .
    \label{eq:induced_kernel_shift}
\end{equation}
Up to a sign, the map from $\kappa$ to the shift is the one of
Eq.~\eqref{eq:injectivity_operator}, so away from the moments where
$r_+-r_-=n$ every admissible variation moves the first missing
kernel, and a variation that moved nothing would be the same-evolution
ambiguity that Theorem~\ref{thm:injectivity} excludes.
Both $\kappa$ and $\gamma^{[1]}$ lie in the commutant
\eqref{eq:commutant_arena}, whose three $\GL(1)$ factors are abelian,
so the commutator survives only in the singlet-gluon block and the two
terms of Eq.~\eqref{eq:induced_kernel_shift} do not reach the same
sectors,
\begin{equation}
\begin{aligned}
    \delta\gamma^{[n+1]}\big|_{\{\Sigma_+,g\}}
    &= \big[\kappa,\gamma^{[1]}\big]\big|_{\{\Sigma_+,g\}}
       - n\,\beta_0\,\kappa\big|_{\{\Sigma_+,g\}},\\[3pt]
    \delta\gamma^{[n+1]}\big|_{i}
    &= -\,n\,\beta_0\,\kappa\big|_{i},
\end{aligned}
\label{eq:induced_shift_sectors}
\end{equation}
where $i$ runs over the three scalar sectors of
Eq.~\eqref{eq:irrep_decomposition} and $\kappa|_{i}$ is the single
function that $\kappa$ carries there. In those sectors the variation is
the running of the coupling alone. The $C$-even scalar is
unconstrained, and the two $C$-odd ones vanish at $z=1$, which is the
condition the kernel itself obeys there, so the family covers every
direction the true kernel can take in these sectors.
The variation samples the unknown $\gamma^{[n+1]}$ along a family
of directions that satisfy the covector conditions automatically, and
Eq.~\eqref{eq:master_shift} converts each member into a shift
$\delta A_n$ in closed form. For the kernel part of the unknown the
two halves of this section are therefore the same estimate seen from
two directions, while the coefficient $\beta_n$ of the running is
left untouched by any scheme transformation. In the language
of theory covariance matrices \cite{NNPDF:2019}, the family provides
the nuisance directions of the evolution.

The minimal version keeps only the moment-independent part, the three
functions of Eq.~\eqref{eq:final_group},
\begin{equation}
\begin{gathered}
    a(\mu) = \kappa_a\, a_s^{\,n}(\mu),
    \qquad
    b(\mu) = 1 + \kappa_b\, a_s^{\,n}(\mu),\\
    \nu(\mu) = 1 + \kappa_\nu\, a_s^{\,n}(\mu),
\end{gathered}
    \label{eq:mhou_family}
\end{equation}
up to corrections of the next order. The natural size of the
parameters follows from Eq.~\eqref{eq:induced_shift_sectors} and the
calibration of Sec.~\ref{sec:mhou_scale},
$|\kappa|\approx|\gamma^{[n+1]}|/(n\beta_0)$, which is about four for
the non-singlet at $z=2$ and $n=1$ and grows toward both ends of the
$z$ axis, logarithmically at large $z$ and through the small-$x$ poles
of the singlet block as $z\to1$.

Like the classification itself, the family is a statement at a fixed
scale, and the evolution does not leave it alone. The relabeling of the
densities shrinks as $\mu$ grows, as Fig.~\ref{fig:orbit} shows, while
what a band measures is the effect of $\delta\gamma^{[n+1]}$
accumulated over the evolution. Equation~\eqref{eq:master_shift}
performs that transport in closed form, with denominators
$n\big(n\pm(R_+-R_-)\big)$ that differ between the two branches and
vanish at the resonant moments of Sec.~\ref{sec:mhou_scale}. One
$\kappa$ therefore produces spreads of different size on $R_+$ and
$R_-$ and at different moments, and the width of a band is not simply
the size of $\kappa$. The recipe is to propagate the
varied kernel through the truncated evolution, for instance through
the coefficients of Eq.~\eqref{eq:frobenius_form}, and to read the
spread, which is of the size of Eq.~\eqref{eq:truncation_scaling} by
construction. Figure~\ref{fig:orbit} shows the failure mode of a
variation outside the classification, with the family taken at $n=1$
for illustration. The three-parameter family is the cheapest option,
while a realistic exercise should sample $\kappa(z)$ from the
constrained family \eqref{eq:mhou_family_general}, for example through
low-order polynomials in $z$ in each block.

\subsection{Relation to scale variation}
\label{sec:scale_variation_relation}

Scale variation fixes the covectors exactly, since
$q^\dagger E_\gamma=q^\dagger$ as Sec.~\ref{sec:covectors} shows, so it
never violates a sum rule. Whether it is admissible in the
sense of Definition~\ref{def:admissible} depends on how the scale is
moved. At leading order the transport between two scales is
\begin{equation}
    E_\gamma(\mu_2,\mu_1)
    = \left(\frac{a_s(\mu_1)}{a_s(\mu_2)}\right)^{\gamma^{[1]}/\beta_0},
    \label{eq:lo_transport}
\end{equation}
so the transformation approaches the identity exactly when the two
couplings approach each other. Moving the scale by a fixed ratio does
that, since $a_s(\mu)/a_s(\xi\mu)\to1$ and the boundary condition
\eqref{eq:asymptotic_identity} holds. Evolving to a scale held fixed
does not, since $a_s(\mu)/a_s(\mu_0)\to0$, which is the caveat that
Sec.~\ref{sec:solution} makes for the schemes that absorb the
evolution.

In the admissible case, expanding Eq.~\eqref{eq:lo_transport} gives
$\kappa=\gamma^{[1]}\ln\xi^2$ at $n=1$, so scale variation by a fixed
ratio is a member of the general family
\eqref{eq:mhou_family_general}, which extends it and does not compete
with it. The commutator drops out because $\kappa$ is proportional to
$\gamma^{[1]}$, so Eq.~\eqref{eq:induced_kernel_shift} leaves
$-\beta_0\ln\xi^2\,\gamma^{[1]}$, the standard shift of the next kernel
when the scale is changed. Only $n=1$ occurs, because a change of scale
starts at order $a_s$ whatever the working order, while the family at
order $a_s^{\,n}$ moves the first missing kernel and leaves the known
ones alone. A factorization scale introduced as an independent scale of
the operator subtraction is a different matter, a genuine deformation
of the scheme and not a motion along the flow, which stays inside the
classification only under the matching condition discussed in
Sec.~\ref{sec:covectors}.

\section{Summary and outlook}
\label{sec:summary}

Let us first summarize the logic of Part~\ref{part:one}. The
symmetries of the kernels act at every Mellin moment and reduce the
transformation to four blocks via Schur's lemma. The sum rules then
act at their own two moments and remove most of what remains, as the
table after Theorem~\ref{thm:classification} records. Finally,
asymptotic freedom, used as a boundary condition, makes the
classification complete, since the map from transformations to kernel
pairs is injective.

Physically, the quark-gluon momentum split and the normalization of
the $C$-even non-singlets are therefore conventions, whereas the
valence content of the hadron and the signs of the real Mellin moments
of every non-singlet distribution are not. Positivity in $x$ space, on
the contrary, is preserved only by a semigroup inside the
classification, whose moment-independent part is given by
Eq.~\eqref{eq:positivity_semigroup}, and even the evolution operator
itself respects the semigroup only at leading order and only upward in
the scale. The same argument decides the polarized case as well, where
the helicity and Soffer bounds tie the unpolarized, helicity and
transversity densities of a flavor together, and a change of scheme
preserves them for every input exactly when its own three kernels obey
the same bounds (Appendix~\ref{sec:soffer}). The sum rules, however,
are never an obstruction by themselves. A transformation with the
right symmetries that fails them is repaired by a normalization that
the sum rules determine, which extends the one-loop momentum
counterterm of Refs.~\cite{Delorme:2025,Delorme:2026} to the number
sum rules and to every order. For collinear densities defined as
integrals of TMDs, in particular, this identifies the scheme that keeps
both the sum rules and the integral relation, namely the unnormalized
scheme times a constant matrix fixed by three numbers
(Sec.~\ref{sec:tmd}).

In Part~\ref{part:two} we then used the same gauge structure to solve
the evolution. Relaxing the boundary condition and gauging the
singlet-gluon connection to zero reduces DGLAP at any order to one
linear second-order equation plus a quadrature
(Sec.~\ref{sec:solution}), in a form that does not change with the
perturbative order. As a consequence, the Frobenius representation of
the solution isolates exactly which coefficients are known at a given
accuracy and the size of the first missing correction
(Sec.~\ref{sec:mhou}), while scheme variation within the classified
transformations estimates the same missing orders and respects the sum
rules exactly (Sec.~\ref{sec:scheme_variation}).

The two parts, moreover, corroborate each other. The linear map that
makes the classification injective is, up to a sign, the map that
turns an admissible scheme variation into a shift of the first missing
kernel, so the uniqueness theorem of Part~\ref{part:one} and the
uncertainty estimate of Part~\ref{part:two} are one operator read in
two directions. The moments where it degenerates are the resonances of
the series solution and the singular denominators of the standard
Mellin-space solution, and yet the evolution operator passes smoothly
through them. Likewise, the computed evolution transports the sum-rule
covectors, so those moments are protected from missing orders too. We
have checked numerically that the series converges at the predicted
rate and behaves as an asymptotic expansion outside its window, that
the dressed expansion of \textsc{Pegasus} reproduces it term by term,
and that the missing-order formulas predict the actual change from NLO
to NNLO. Furthermore, the leading-order kernels lie inside the
positivity semigroup, which recovers the known survival of positivity
and of the Soffer bound under leading-order evolution
\cite{Barone:1997,Vogelsang:1997}. In the end, both parts rest on the
same observation, namely that scheme transformations are the gauge
group of DGLAP evolution.

We conclude with some directions for future work. The most useful
next step, in our view, is to place the schemes that are actually used
in practice within the classification, along the lines of
Sec.~\ref{sec:tmd}. The DIS scheme and the schemes collected in
Refs.~\cite{Delorme:2025,Delorme:2026} are natural cases, and for each
of them the question is the same one we asked of the cutoff scheme,
namely which sum rules it keeps and what normalization restores the
others. Flavor thresholds deserve a dedicated treatment as well, since
crossing a threshold changes $N_f$ and hence the group, and matching
is itself a scheme transformation at the boundary. Once these pieces
are in place, a global fit that varies the scheme within the
classified transformations would show whether a sum-rule-preserving
uncertainty band differs in practice from the usual scale-variation
band.

Moreover, little of what we did depends on the densities being
spacelike. The closest case is that of the fragmentation functions,
whose evolution has the same flavor structure and the same commutant.
Since the boundary condition and the completeness argument only use
asymptotic freedom, and the solution of Part~\ref{part:two} needs only
the two-by-two singlet block, everything carries over with the
timelike kernels in place of the spacelike ones. However, the sum
rules are not the same. Fragmentation functions obey only the momentum
sum rule, summed over hadrons, because the first moment is a
multiplicity that is not finite. Nothing is therefore frozen at $z=1$,
and the classified group is the momentum row of the table after
Theorem~\ref{thm:classification}, with five free functions in place of
three. The integral relation between TMD and collinear fragmentation
functions holds as well, with logarithms of the momentum fraction in
the matching coefficients \cite{Gonzalez-Hernandez:2023}, so the
normalization of Sec.~\ref{sec:final_cut} is needed only at $z=2$, and
it then involves two numbers in place of three.

Finally, nothing in the construction is specific to parton densities
at all. Any renormalization-group equation that is linear and
homogeneous, whatever the convolution that a suitable transform turns
into a product, is a connection on the scale axis whose finite
renormalizations act by the same transformation law. Wherever a
conserved quantity gives a covector annihilated by the anomalous
dimension and asymptotic freedom gives a boundary condition, the
classification and the uniqueness theorem therefore carry over, with
the symmetries of that problem in place of charge conjugation and
flavor symmetry, and so does the solution of Part~\ref{part:two} for
every two-dimensional mixing block.

\section*{Acknowledgments}

The author thanks Ted Rogers for useful discussions. The author is supported by the DOE, Office of Science, Office of Nuclear Physics, Early Career Program under contract No. DE-SC0025881.

\appendix

\section{Positivity with helicity and transversity}
\label{sec:soffer}

The positivity argument of Sec.~\ref{sec:positivity} extends to the
polarized densities, where the physical constraints tie three
distributions together. For a quark of one flavor the unpolarized
density $q$, the helicity density $\Delta q$ and the transversity
density $\delta q$ obey
\begin{equation}
    q - \Delta q \geqslant 0,
    \qquad
    q + \Delta q \geqslant 0,
    \qquad
    2\,|\delta q| \leqslant q + \Delta q ,
    \label{eq:soffer_bounds}
\end{equation}
at every momentum fraction. The first two are the helicity bound
$|\Delta q|\leqslant q$ and the third is the Soffer bound
\cite{Soffer:1995}. The asymmetry of the third one is physical, since
transversity flips the helicity of the quark and of the hadron
together and is therefore bounded by the configuration in which the
two helicities are aligned, whose weight is $q+\Delta q$.

The three densities evolve with three different kernels and carry
three separate scheme freedoms, so a change of scheme acts through
three $x$-space kernels,
\begin{equation}
    q' = K\otimes q,
    \qquad
    \Delta q' = \Delta K\otimes\Delta q,
    \qquad
    \delta q' = \delta K\otimes\delta q ,
    \label{eq:polarized_map}
\end{equation}
each a function plus a multiple of $\delta(1-x)$ as in
Theorem~\ref{thm:positivity}. The combinations that appear in
Eq.~\eqref{eq:soffer_bounds} mix under this map,
\begin{equation}
    q'\pm\Delta q'
    = \tfrac12(K\pm\Delta K)\otimes(q+\Delta q)
    + \tfrac12(K\mp\Delta K)\otimes(q-\Delta q) ,
    \label{eq:helicity_mix}
\end{equation}
so the helicity-conserving kernel $(K+\Delta K)/2$ and the
helicity-flip kernel $(K-\Delta K)/2$ are the objects that act.

\begin{theorem}[Polarized positivity]
\label{thm:soffer}
The map \eqref{eq:polarized_map} sends every triple obeying
Eq.~\eqref{eq:soffer_bounds} into a triple obeying
Eq.~\eqref{eq:soffer_bounds} exactly when the kernels themselves obey
the same bounds,
\begin{equation}
    K - \Delta K \geqslant 0,
    \qquad
    K + \Delta K \geqslant 0,
    \qquad
    2\,|\delta K| \leqslant K + \Delta K ,
    \label{eq:soffer_conditions}
\end{equation}
at every momentum fraction.
\end{theorem}

\noindent
Necessity follows as in Sec.~\ref{sec:positivity}. An input
concentrated at one momentum fraction isolates the kernels at one
ratio, so it is enough to treat the map with each kernel replaced by
its value there. A maximally polarized input, $q=\Delta q$ with
$\delta q=0$, makes Eq.~\eqref{eq:helicity_mix} read
$q'-\Delta q'=\tfrac12(K-\Delta K)(q+\Delta q)$ and forces
$K-\Delta K\geqslant0$, and the oppositely polarized input
$q=-\Delta q$ forces $K+\Delta K\geqslant0$ the same way. Taking
$q=\Delta q$ again and saturating the Soffer bound,
$2\delta q=q+\Delta q$, gives $2|\delta q'|=|\delta K|\,(q+\Delta q)$
against $q'+\Delta q'=\tfrac12(K+\Delta K)(q+\Delta q)$, which is the
third condition. Sufficiency is the chain
\begin{equation}
\begin{aligned}
    2|\delta q'| &\leqslant 2|\delta K|\otimes|\delta q|
    \leqslant |\delta K|\otimes(q+\Delta q)\\
    &\leqslant \tfrac12(K+\Delta K)\otimes(q+\Delta q)
    \leqslant q'+\Delta q' ,
\end{aligned}
    \label{eq:soffer_chain}
\end{equation}
whose last step drops the second term of
Eq.~\eqref{eq:helicity_mix}, nonnegative because both of its factors
are. For the singlet the first two conditions apply entry by entry to
the pair $(q,\Delta q)$ including the gluon, while transversity has no
gluon partner at leading twist and enters only through the quark
entries.

The leading-order kernels satisfy Eq.~\eqref{eq:soffer_conditions}.
We write $P^{[k]}$ for the $x$-space kernels of the expansion of
Sec.~\ref{sec:mellin}, so that $\gamma^{[k]}=\Mellin\{P^{[k]}\}$ and
the leading one is $P^{[1]}$. The plus-distribution and $\delta(1-x)$
parts cancel in every combination, and what remains is a nonnegative
function,
\begin{equation}
\begin{gathered}
    P^{[1]}_{qq}+\Delta P^{[1]}_{qq}-2\,\delta P^{[1]}_{qq}
    = 4C_F\,(1-x),\\
    P^{[1]}_{qq}-\Delta P^{[1]}_{qq} = 0,
    \qquad
    P^{[1]}_{gg}-\Delta P^{[1]}_{gg} = 4C_A\,\frac{(1-x)^3}{x},\\
    P^{[1]}_{gq}-\Delta P^{[1]}_{gq} = 4C_F\,\frac{(1-x)^2}{x},\\
    P^{[1]}_{qg}-\big|\Delta P^{[1]}_{qg}\big|
    = 8N_fT_R\,\min(x,1-x)^2 ,
\end{gathered}
\label{eq:lo_soffer_check}
\end{equation}
with $C_F$, $C_A$ and $T_R$ the usual color factors and every sum
$P^{[1]}+\Delta P^{[1]}$ nonnegative as well. The leading-order
evolution operator over any upward step therefore lies in this
semigroup, which recovers the known result that both bounds survive
leading-order evolution \cite{Barone:1997,Vogelsang:1997}, and beyond
leading order Eq.~\eqref{eq:soffer_conditions} decides scheme by
scheme. A transformation acting on transversity alone, with
$K=\Delta K=\delta(1-x)$, preserves the bounds for every input only if
it is a pointwise rescaling, $\delta K$ proportional to $\delta(1-x)$
with a factor of modulus at most one, since any support away from the
endpoint violates the saturated bound for an input that grows toward
small $x$. This is the transversity twin of the moment-independent
case of Eq.~\eqref{eq:positivity_semigroup}.

\section{Relation to the dressed expansion of the evolution operator}
\label{sec:dressed}

Section~\ref{sec:frobenius} compared the Frobenius series with the
dressed expansion of the evolution operator used in
\textsc{Pegasus} \cite{Vogt:2005}. Here we make the identification
term by term.

Both constructions solve the same equation. At fixed $z$, trading
$\ln\mu^2$ for the coupling puts the singlet evolution in the form
\begin{equation}
\begin{gathered}
    \frac{\dd \overline f}{\dd a_s}
    = -\frac{1}{a_s}\,\Gamma(a_s)\,\overline f,\\
    \Gamma(a_s) = \sum_{k\geqslant0} a_s^k\, \Gamma_k,
    \qquad
    \Gamma_0 = \frac{\gamma^{[1]}}{\beta_0},
\end{gathered}
    \label{eq:a_equation}
\end{equation}
where the matrices $\Gamma_k$ collect the kernels and the expanded $\beta$
function of the chosen truncation. At NLO, for instance,
$\Gamma_k = (-\beta_1/\beta_0)^{k-1}\big(\gamma^{[2]}/\beta_0
-(\beta_1/\beta_0)\,\Gamma_0\big)$ for $k\geqslant1$. The dressed expansion
writes the solution as the leading-order one corrected on both sides
by a power series of matrices,
\begin{equation}
\begin{gathered}
    \overline f(a_s) = U(a_s)\;
    e^{-\Gamma_0\ln(a_s/a_0)}\;
    U(a_0)^{-1}\,\overline f(a_0),\\
    U(a_s) = \Id + \sum_{k\geqslant1} a_s^k\, U_k ,
\end{gathered}
    \label{eq:dressed_ansatz}
\end{equation}
and inserting the ansatz into Eq.~\eqref{eq:a_equation} gives the
commutator recursion
\begin{equation}
    k\,U_k + [\Gamma_0,\,U_k] = -\sum_{i=1}^{k} \Gamma_i\,U_{k-i} .
    \label{eq:U_recursion}
\end{equation}
Writing $\Pi_\pm$ for the projectors onto the eigenvalues $r_\pm$ of
$\Gamma_0$, the recursion inverts component by component with the
denominators
\begin{equation}
\begin{gathered}
    \Pi_a\,U_k\,\Pi_b
    = \frac{1}{k + r_a - r_b}\;
    \Pi_a\Big(-\sum_{i=1}^{k} \Gamma_i\,U_{k-i}\Big)\Pi_b ,\\
    a,b \in \{+,-\},
\end{gathered}
    \label{eq:U_projector}
\end{equation}
which are singular exactly when the eigenvalue gap $r_+-r_-$ equals
the integer $k$.

The Frobenius picture starts instead from the two vector solutions of
Eq.~\eqref{eq:a_equation},
\begin{equation}
    v_\pm(a_s) = a_s^{-r_\pm}\sum_{k\geqslant0} c_k^{(\pm)}\,a_s^k,
    \qquad
    c_0^{(\pm)} = w_\pm ,
    \label{eq:frobenius_vector}
\end{equation}
with $w_\pm$ the eigenvectors of $\Gamma_0$. The two constructions generate
the same expansion. Applying Eq.~\eqref{eq:U_recursion} to $w_\pm$ and
using $\Gamma_0 w_\pm = r_\pm w_\pm$ reproduces exactly the recursion
satisfied by the $c_k^{(\pm)}$, with the same starting value, so by
induction
\begin{equation}
    c_k^{(\pm)} = U_k\, w_\pm .
    \label{eq:UW_identity}
\end{equation}
The columns of the dressing matrices in the leading-order eigenbasis
are the coefficients of the two Frobenius series, and the operator of
Eq.~\eqref{eq:dressed_ansatz} is the ratio of the fundamental matrices
built from Eq.~\eqref{eq:frobenius_vector}.

The scalar series of Sec.~\ref{sec:frobenius} stores the same
information more economically, and the correspondence is explicit.
Undoing the substitution of Sec.~\ref{sec:riccati} identifies the
scalar solution with the singlet component of a vector solution times
the $\omega_q$ factor of Eq.~\eqref{eq:K0_explicit}, which splits into
a power and an analytic part,
$\omega_q = a_s^{\gamma^{[1]}_{qq}/\beta_0}\,\Omega_q(a_s)$ with
$\Omega_q = (1+\beta_1 a_s/\beta_0)^{\tau_q}$ and
$\tau_q = (\beta_0\gamma^{[2]}_{qq}
-\beta_1\gamma^{[1]}_{qq})/(\beta_0\beta_1)$ at NLO. The power
converts the eigenvalue into the index of
Eq.~\eqref{eq:frobenius_indices}, with the two solutions crossing
over, and the coefficients follow by expanding the product,
\begin{equation}
    u^{(\pm)}_{qq} = \omega_q\,\big(v_\mp\big)_q ,
    \quad
    A_m^{(\pm)} = \frac{1}{(w_\mp)_q}\sum_{j=0}^{m}
    \Omega_{q,j}\,\big(U_{m-j}\,w_\mp\big)_q ,
    \label{eq:Am_dictionary}
\end{equation}
where $(\,)_q$ denotes the singlet component, $\Omega_{q,j}$ are the
Taylor coefficients of $\Omega_q$, and the normalization enforces $A_0=1$.
The row-wise shift between the indices \eqref{eq:frobenius_indices}
and the eigenvalues $r_\pm$ drops out of every difference, so the
resonance condition is common to the two pictures. Merging the two
components into one second-order equation multiplies their
denominators, $k$ and $k\pm(r_+-r_-)$, into the indicial factor
$F(R_\pm+k)=k\,\big(k\pm(R_+-R_-)\big)$ of
Eq.~\eqref{eq:frobenius_recursion}.

The difference between the methods is therefore not in the terms but
in the policy. \textsc{Pegasus} truncates $U$ at the order of the
kernels, with variants that differ beyond it, while the Frobenius
series is resummed, with the radius and the error anatomy discussed in
the main text. Since the terms coincide, there is no difference in
convergence to speak of. In fact, truncating the two expansions at
the same order gives the same evolution operator, since by
Eq.~\eqref{eq:UW_identity} the truncated dressing applied to the
leading-order solutions is the truncated fundamental matrix, and the
factors between them cancel in the operator. Neither policy is
systematically closer to the true evolution either, because the equation being solved is
itself truncated and the difference between resumming and stopping at
the kernel order is of the size of the missing kernels, which is
exactly the uncertainty quantified in Sec.~\ref{sec:mhou}. We have confirmed the identification numerically with
the setup of Sec.~\ref{sec:frobenius} for $z=2$ and $z=5$ as illustrative examples. Iterating the dressing to all
orders reproduces the exact solution of the truncated-kernel equation
to the accuracy of the integrator, Eq.~\eqref{eq:UW_identity} holds to
numerical accuracy, and $\|U_2\|$ develops a simple pole at the lower
$r_+-r_-=2$ resonance, $z\simeq1.49$, of Fig.~\ref{fig:radius}.

\bibliographystyle{apsrev4-2-inspire}
\bibliography{bibliography}

\end{document}